\documentclass[journal=ancac3,manuscript=article]{achemso}
\usepackage[version=3]{mhchem} 
\usepackage{graphicx}
\usepackage{color,soul}
\usepackage{hyperref}
\usepackage{comment}
\usepackage{gensymb}
\usepackage{siunitx}
\usepackage{natmove}
\usepackage{tabularx}

\newcommand{\mose} {MoSe$_2$}
\newcommand{\nbse}{NbSe$_2$}

\newcommand{\mos}{MoS$_2$}
\newcommand{\xt}{X$^-$}
\newcommand{\xo}{X$^0$}

\author{Jaydeep Joshi}
\affiliation{Department of Physics and Astronomy, George Mason University, Fairfax, Virginia 22030, United States}
\alsoaffiliation{Quantum Materials Center, George Mason University, Fairfax, Virginia 22030, United States}

\author{Tong Zhou}
\affiliation{Department of Physics, University at Buffalo, Buffalo, New York 14260, United States}

\author{Sergiy Krylyuk}
\affiliation{Materials Science and Engineering Division, National Institute of Standards and Technology, Gaithersburg, Maryland 20899, United States}

\author{Albert V. Davydov}
\affiliation{Materials Science and Engineering Division, National Institute of Standards and Technology, Gaithersburg, Maryland 20899, United States}

\author{Igor Zutic}
\affiliation{Department of Physics, University at Buffalo, Buffalo, New York 14260, United States}

\author{Patrick M. Vora}
\affiliation{Department of Physics and Astronomy, George Mason University, Fairfax, Virginia 22030, United States}
\alsoaffiliation{Quantum Materials Center, George Mason University, Fairfax, Virginia 22030, United States}
\email{pvora@gmu.edu}

\title{Localized Excitons in \nbse{} - \mose{} Heterostructures}

\keywords{excitons, van der Waals heterostructures, transition metal dichalcogenides, photoluminescence, density functional theory}

\begin{document}

\begin{abstract}
Neutral and charged excitons (trions) in atomically-thin materials offer important capabilities for photonics, from ultrafast photodetectors to highly-efficient light-emitting diodes and lasers. Recent studies of van der Waals (vdW) heterostructures comprised of dissimilar monolayer materials have uncovered a wealth of optical phenomena that are predominantly governed by interlayer interactions. Here, we examine the optical properties in \nbse{} - \mose{} vdW heterostructures, which provide an important model system to study metal-semiconductor interfaces, a common element in optoelectronics. Through low-temperature photoluminescence (PL) microscopy we discover a sharp emission feature, L1, that is localized at the \nbse{}-capped regions of \mose{}. L1 is observed at energies below the commonly-studied \mose{} excitons and trions, and exhibits temperature- and power-dependent PL consistent with exciton localization in a confining potential. Remarkably, L1 is very robust not just in different samples, but also under a variety of fabrication processes. Using first-principles calculations we reveal that the confinement potential required for exciton localization naturally arises from the in-plane band bending due to the changes in the electron affinity between pristine \mose{} and \nbse{} - \mose{} heterostructure. We discuss the implications of our studies for atomically-thin optoelectronics devices with atomically-sharp interfaces and tunable electronic structures.
\end{abstract}

\section{Introduction}

Semiconductor heterostructures are key to modern technology and devices such as transistors, lasers, solar cells, and light-emitting diodes.\cite{Alferov2002} However, the growth constraints of lattice matching between different semiconductors significantly limit not only the number of semiconductor heterostructures, but also their ultimate performance and functionality. These constraints can be overcome with the growing number of two-dimensional (2D) crystals in which different atomic monolayers (MLs) are coupled by weak van der Waals (vdW) forces. Stacking 2D crystals leads to vdW heterostructures which are immune from growth rules, capable of forming atomically-sharp interfaces, and also transform materials physics through proximity effects.\cite{Zutic2019} In fact, even transition metal dichalcogenides (TMDs), a small subset of these 2D crystals, embody a wide range of materials properties, including being semiconducting, metallic, magnetic, superconducting, or topologically nontrivial.\cite{Choi2017,Manzeli2017,Mak2016,Guguchia2018} The electronic structure in TMDs combines time-reversal symmetry and broken inversion symmetry, which has important implications for optical and transport properties. Semiconducting ML TMDs have direct band gaps,\cite{Mak2010} very large excitonic binding energies (up to ∼0.5 eV),\cite{Chernikov2014} efficient light emission,\cite{Wang2017a,Mueller2018} and a characteristic coupling between the valleys K and K' and the carrier spin.\cite{Xu2014}

An important opportunity for vdW heterostructures lies in the potential to interface layered materials with dissimilar properties, dimensionality and lattices, a capability that could enable complex device functionalities impossible with conventional materials.\cite{Jariwala2017,Liu2016b} Unfortunately, TMD-based devices with common metals often lead to highly-resistive contacts\cite{Liu2012,Das2013a,Larentis2012,Das2013} and limit the possibility to effectively combine different materials including the opportunity to employ desirable proximity effects.\cite{Zutic2019} Furthermore, such metals are typically also responsible for a rapid decay of charge carriers from the semiconductor resulting in significant quenching of photoluminescence (PL) emission,\cite{Bhanu2014,Froehlicher2018,Zhang2019} constraining their use in TMD-based optoelectronic devices.

A promising direction to solve these obstacles relies on using atomically-thin metals to create low-resistance and highly transparent vdW contacts. Particularly interesting are Nb-based metallic TMDs which offer low, p-type Schottky barrier with many 2D semiconductors in addition to exhibiting superconductivity.\cite{Guan2017b,Shin2018,Dvir2018,Huang2018,Lv2018a}  This has enabled further developments in fabricating ultrathin devices, where drastic changes in the electronic landscape of the \nbse{} contact can modulate the channel carrier properties of the transistor.\cite{Huang2018,Yabuki2016,Lee2019,Kim2017d} On the other hand, extensive investigations to understand interfacial contact properties and nearby dielectrics in \mose{}-based metal-semiconductor heterostructures has paved the way to engineer efficient optoelectronic devices.\cite{Pan2016,Liu2018b,Cakir2014,Huang2017a} However, despite this progress in the studies of 2D metal-semiconductor heterostructures, important uncertainties in their optical properties and the role of excitons remain. Fermi level pinning at metal-semiconductor interfaces can induce mid-gap states in the semiconductor, thereby enabling new radiative channels for exciton recombination.\cite{Liu2016c,Monch1999,Heine1965,Sajjad2019,Lu2014}
Semiconducting ML TMDs pose additional challenges with the influence of surrounding dielectric media, where 2D metals can drastically change the exciton dynamics as well as the ability to effectively screen electrons in the semiconductor.\cite{Scharf2019}

Motivated by the promising properties of \nbse-based contacts with semiconducting TMDs as well their largely unexplored optical response, we perform hyperspectral and temperature and power-dependent PL spectroscopy of heterostructures composed of bulk-\nbse{} and ML-\mose{}. Remarkably, even under very different assembly conditions in 10 samples, we reveal a robust novel spectral feature, L1, which is unique to metal-semiconductor interface. L1 consistently appears between 1.59 eV - 1.6 eV, approximately 25-30 meV below the \mose{} trion and has stronger temperature changes than other excitonic emissions, while it exhibits a sub-linear excitation-power dependence. We observe that for a particular fabrication procedure of \nbse{} - \mose{} heterostructures there is a splitting of the exciton and trion, which may signal the relative difference in the quality of the interfacial contact across the studied samples. We explain L1 as a localized exciton state in deep potential well traps provided by an in-plane band-offset between the bare \mose{} and its region capped by \nbse{}. The implications of L1 for emerging vdW-based optoelectronic devices are discussed.

\section{Results and Discussion}

\begin{figure}[h]
    \centering
    \includegraphics[width=6.5in]{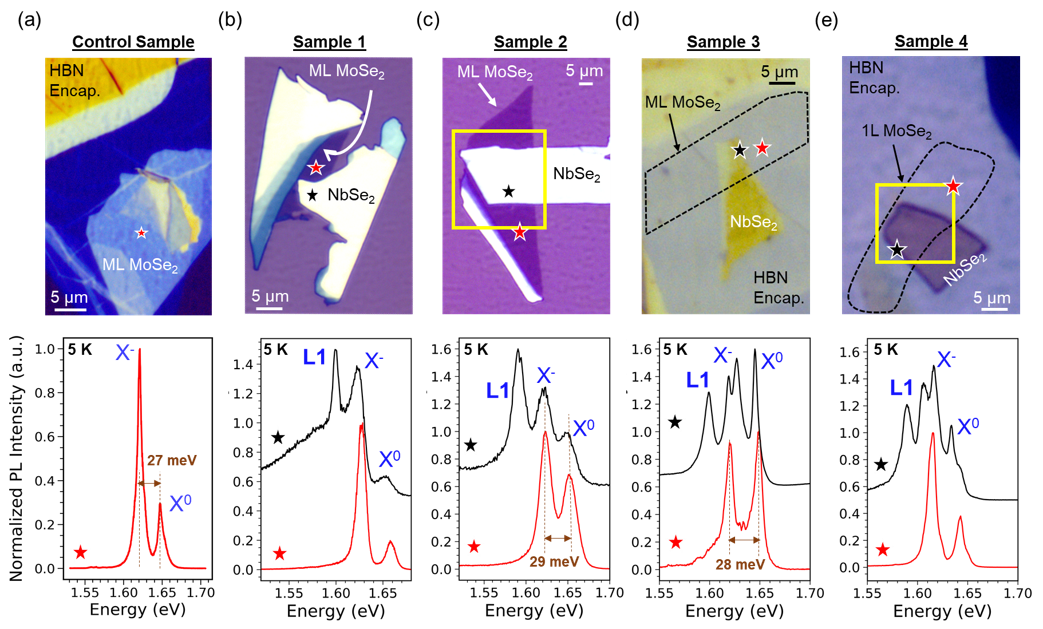}
    \caption{White light images of (a) fully encapsulated ML-\mose{} Control sample and (b-e) Samples 1, 2, 3 and 4 comprising of \nbse{} - \mose{} vdW heterostructures. Samples 1 and 2 are fabricated on Si/SiO$_2$ substrate, whereas Samples 3 and 4 are fully encapsulated with few-layer hexagonal boron nitride (hBN). The black dotted line in (d) and (e) outlines the ML-\mose{} region. Beneath each image are low-temperature PL spectra taken on (black) and off (red) the \nbse{} - \mose{} interface. These locations are denoted by red and black stars in the white light image. The yellow boxes mark the regions scanned in hyperspectral PL maps. In addition to the \xo\ and \xt\ emission observed in Samples 1-4, a new PL feature referred to as L1 appears at $\approx$ 1.6 eV in all heterostructures.} 
    \label{allsamp}
\end{figure}

\subsection{Optical characterization of vdW heterostructures}

White light images of a subset of \nbse{} - \mose{} vdW heterostructures and a Control sample are presented in Figure \ref{allsamp}. The Control sample (Figure \ref{allsamp}a) consists of  ML-\mose{} fully encapsulated with few-layer hexagonal boron nitride (hBN). Samples 1 and 2 are assembled directly on a Si/SiO$_2$ substrate without encapsulation, as shown in Figures \ref{allsamp}b and \ref{allsamp}c.  Samples 3 and 4 (Figures \ref{allsamp}d and \ref{allsamp}e) are both fully encapsulated with few-layer hBN, with ML-\mose{} regions outlined by dashed black lines. Below each image are low-temperature (5 K) PL spectra showing emissions on (black star) and off (red star) the \nbse{} - \mose{} interface shown in the corresponding image. All samples exhibit PL features close to 1.65 eV and 1.62 eV attributed to neutral exciton (\xo) and negatively-charged trion (\xt) emissions in ML-\mose, respectively.\cite{Ross2013} We expect the trion to be \xt\ and not X$^+$ owing to the formation of Se vacancies and anti-sites during synthesis and the tendency of \mose{} to be inherently n-type.\cite{Edelberg2019} The observed \xt\ to \xo\ binding energies ($\approx$ 27-30 meV) and linewidths of these transitions are ($\approx$ 4-6 meV for Samples 3 and 4, $\approx$ 6-9 meV for Samples 1 and 2) consistent with prior measurements,\cite{Lu_2019,Ajayi2017,Wierzbowski2017,Jadczak2017a,Ross2013} and their variation likely due to dielectric tuning of the Coulomb interaction. \cite{Scharf2019,VanTuan2018a,Florian2018} It is also important to point out that relative intensity of \xt\ peak to the \xo\ is much higher in majority of our samples (Control and Samples 1, 2 and 4 in Figure \ref{allsamp} as well as six other samples included in Supporting Information (S1)), also indicating an intrinsically n-doped \mose{} source, however, additional charge doping mechanism from substrate or polymer-residue interaction may also electronically dope the monolayer.

The most obvious difference between on (black) and off (red) spectra is the emergence of a new PL peak close to $\approx$ 1.6 eV, about 25-30 meV lower than the \xt\ peak.  The intensity and linewidth of this peak, hereafter referred to as L1, varies between samples, however, its energy with respect to the \xt\ peak remains unchanged and its emission localized to the \nbse{} - \mose{} interface. L1 is also surprisingly robust to the quality of the interfacial contact between the two TMDs, as demonstrated by its presence in 6 additional heterostrucutres included in Supporting Information (S1) fabricated in a variety of stacking configurations and post-assembly cleaning procedures (see Methods section). Furthermore, we find that  the \xo\ and \xt\ PL peaks split in hBN-encapsulated Samples 3 and 4 (Figure \ref{allsamp}d and \ref{allsamp}e) leading to a 5 peak structure at the \nbse{} - \mose{} interface. We believe this to be a consequence of intra-layer moir\'{e} exciton at interface formed between \nbse{} and \mose, similar to the observations in \mose{} - \mos\ heterostructure (discussed in Supporting Information - S2).\cite{Zhang2018b} 

\begin{figure}[h]
    \centering
    \includegraphics[width=6.5in]{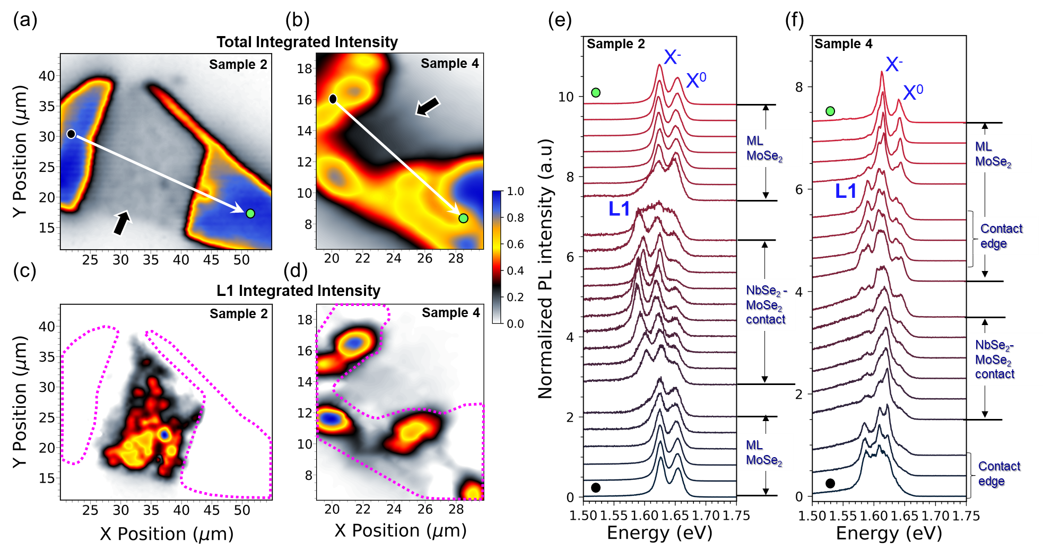}
    \caption{Low-temperature hyperspectral maps of the total integrated PL intensity from 1.5 eV - 1.7 eV for (a) Sample 2 and (b) Sample 4, and integrated PL intensity of L1 peak for (c) Samples 2 and (d) Sample 4. The regions scanned correspond to the yellow boxes in Figure~\ref{allsamp}. White arrow in panels (a) and (b) corresponds the path of the linecuts examined in (e) and (f). The magenta dashed outlines in panels (c) and (d) mark bare ML-\mose{}. PL spectra extracted along the linecuts in (a) and (b) plotted for (e) Sample 2 and (f) Sample 4. L1 is localized along the \nbse{} - \mose{} perimeter in Sample 4, while it appears across the \nbse{} - \mose{} interface in Sample 2. The black and green dots refer to the start and end positions of the linecuts, respectively.}
    \label{linecuts}
\end{figure}

Hyperspectral PL measurements allow us to understand the spatial origin of L1 and determine whether vdW coupling between \nbse{} and \mose{} has any contribution. Figure \ref{linecuts}a and \ref{linecuts}b here show the spatially-mapped integrated PL intensity acquired at 5 K for Sample 2 and Sample 4 (hBN-encapsulated), respectively across area marked by yellow boxes in Figures \ref{allsamp}c and \ref{allsamp}e. The black arrows indicate regions of \nbse{} - \mose{} overlap. We first examine the variation in integrated PL intensity (total sum of \xo, \xt and defect emission), where, compared to the bright emission observed from the uncapped region, a reduction by a factor of five is seen at the interfacial contact between \nbse{} and \mose{} flakes. However, the fact that any PL signals are recorded from the \nbse{}-capped regions suggests that non-radiative relaxation into \nbse{} is a weaker process than might be expected. The loss in the total PL intensity at the interface could be a result of bulk-\nbse{} reflecting the incoming laser power and reducing the excitation power density at the monolayer underneath, as well as absorbing the outgoing \mose{} PL, all in addition to the unaccounted interference effects.\cite{Lien2015a} Figure \ref{linecuts}c and \ref{linecuts}d show variation in the integrated intensity of L1 (method outlined in Supporting Information - S3) measured across the same area. These maps highlight an important difference between the behavior of L1 in encapsulated and unencapsulated samples. As seen in Figure \ref{linecuts}c, for Sample 2 fabricated without hBN-encapsulation, L1 is present over the majority of the \nbse{} - \mose{} interface. This in contrast to the hBN-encapsulated Sample 4, where L1 is more prominent over the perimeter of the \nbse{} - \mose{} region, as seen in Figure \ref{linecuts}d. The impact of interface quality can be made more apparent by examining PL plots of spectra extracted along a line-cut, depicted by the white arrow that extends over the interface from the pristine \mose{} regions in Figures \ref{linecuts}a and \ref{linecuts}b. The green and black dots represent the starting and ending point of the line-cut, respectively. It is clear from Figure \ref{linecuts}e (Sample 2) that as we move from ML-\mose{} to the region in contact with \nbse{}, L1 appears at about 30 meV below the \xt\ peak. In contrast to Sample 2, Figure \ref{linecuts}f shows that L1 is more pronounced along the edges surrounding the heterostructure in Sample 4 than at the interface. Additionally, for both samples, we see clear broadening of the \xo\ and \xt\ peaks at the interface compared to the bare \mose{} region. This is typical for metal-semiconductor junctions where rapid decay of charge carriers can result in reduced exciton lifetimes and inhomogeneously broadened PL linewidths. Also in case of Sample 2, the intensity of L1 peak over the interface surpasses the \xo\ and \xt peaks, revealing insights into favourable routes for low-temperature exciton recombination at the \nbse{} - \mose{} interface. We believe the absence of L1 at the interfacial region in Sample 4 is a consequence of uniform coupling between the \nbse{} and \mose{} layers as a result of intermediate ‘nano-squeegeeing’ during sample fabrication (discussed later). Both Samples 2 and 4 were vacuum annealed, which has shown in some cases to improve van der Waals coupling between TMDs, as well as a method to redistribute polymer residue to form larger bubbles.\cite{Khestanova2016,Frisenda2018a,Onodera2019,Tyurnina2019} This can further result in variable contact between two TMDs, forming randomly distributed pockets with better coupling than the surrounding region. In both cases, the prevalence of L1 at and surrounding the interface constitutes a need to theoretically investigate interlayer interactions between \nbse{} and \mose{} and propose a model that explains this unconventional behavior.


\subsection{\nbse{} - \mose{} contact: First-principles calculations}
 
In this section, we focus on understanding and distinguishing the origin of the L1 peak from previously studied, low-energy optical pathways in ML-\mose. We begin by presenting a brief survey of explorations to engineer \mose{} exciton states, multi-exciton \mose{} states, and excitons in \mose{}-based vdW heterostructures, and discuss how L1 defies explanation by these prior works. The controlled application of strain to ML-\mose{} was found to induce localized quantum emitters with ultrasharp PL linewidths over a wide range of energies below \xo\ and \xt.\cite{Branny2016} While an attractive explanation, L1 exhibits a PL linewidth that is much broader than a strain-induced quantum emitter and a consistent emission energy of $\approx$ 1.6 eV, disagreeing with discrete quantum-dot picture. On the other hand, a broad low-energy feature (between 1.5 eV and 1.6 eV) appears in gate-dependent PL measurement of pristine ML-\mose, but is associated with impurity trapped excitons with linewidths an order of magnitude larger than \xo{}, \xt{} and L1.\cite{Ross2013} Another possible explanation may be that L1 arises from dark exciton states, however, the dark exciton in \mose{} not only  occurs at energies higher than exciton ground state,\cite{Deilmann2017} but has recently been observed 1 meV above \xo{},\cite{Lu_2019} in clear disagreement with the emission energy of L1 at 60 meV below \xo{}. Strain-dependent measurement on ML-\mose{} has revealed a uniform red shift in the \xo{} and \xt{} peak energies, without the appearance of a new low-energy excitation.\cite{Aas2018} Measurements to probe higher-order exciton complexes such as neutral or charged biexcitons have revealed features at energies lower than the \xo, however their binding energies do not match the spectral position of L1.\cite{Hao2017a} Finally, emergence of interlayer valley excitons in \mose-based vdW heterostructures occur at  about $\approx$ 1.4 eV much lower than our observation of L1.\cite{Hanbicki2018} Thus, the spectral and spatial properties of L1 are unique to the \nbse{} - \mose{} geometry discussed in our manuscript and demands a thorough investigation of its origin pertaining to the interaction between metallic and semiconducting TMDs.

\begin{figure}[h]
    \centering
    \includegraphics[width=6.5in]{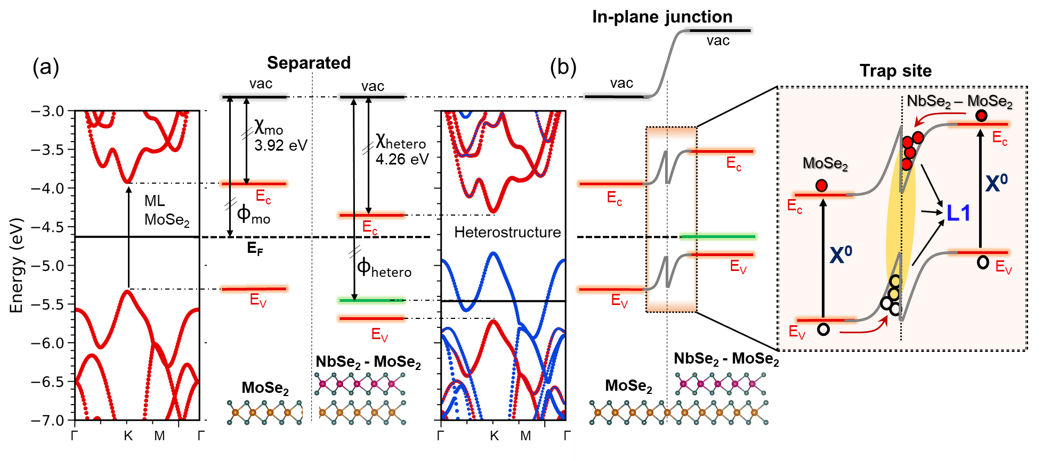}
    \caption{(a) A schematic of changes in the work function ($\Phi$) and electron affinity ($\chi$) between \nbse{} - \mose{} contact and pristine \mose{} traced from their respective DFT calculated bandstructures. Red bands derive from \mose, while blue bands derive from \nbse. The Y-axis for both bandstructures is scaled such that the zero energy refers to the vacuum energy. (b) Discontinuous band bending allows for formation of potential traps at the in-plane junction as the Fermi Level ($E_F$) matches between the two region. A zoom in at the junction shows confinement of  electrons and holes in the valence band (VB) and conduction band (CB) of capped and uncapped \mose. Emissions from the bare and capped \mose{} regions where in addition to the excitons, possibilities of low-energy emissions from recombination of trapped electrons and holes emerge across the interface (yellow).}
    \label{traps}
\end{figure}

In order to qualitatively assess the effects of interaction between \nbse{} and \mose, we perform first-principles calculations of the \nbse{} - \mose{} heterostructure. Density functional theory (DFT) calculations are presented in Figure \ref{traps} for the pristine and heterostructure system. Here, Figure \ref{traps}a illustrates the ML-\mose{} bandstructure (left plot - red bands) which hosts a direct band gap at the $K$ point in agreement with previous calculations.\cite{Island2016,Kang2013a}. The calculated electron affinity of ML-\mose{} is $\chi_{Mo}$ = 3.92 eV, and the Fermi level ($E_F$) and work function ($\Phi_{Mo}$) are set mid-gap based on the assumption of an undoped monolayer in agreement with recent work.\cite{Zhang2019a} Bandstructure calculations for a ML-\nbse{} - ML-\mose{} heterostructure are also presented in Figure \ref{traps}a (right plot) where the blue (red) bands derive from \nbse{} (\mose{}) orbitals, respectively, and qualitatively match prior modeling efforts of this system.\cite{Lv2018a} We note that although the top \nbse{} flake is bulk ((60-100) nm) in our case (Supporting Information - S4), our use of ML-\nbse{} in these calculations is sufficient to capture the essential physics of the system as van der Waals interactions only get weaker for subsequent layers. Also included in the Supporting Information (S4) is the electronic bandstructure for ML-\nbse{} which is metallic with a calculated workfunction value of $\Phi_{Nb}$ = 5.56 eV, in agreement with previous work.\cite{Shimada1994} Upon first examination, the \mose{} bandstructure is qualitatively unchanged by the presence of \nbse{}. The \mose{} bandgap ($E_g$) remains the same and there is no band-hybridization between the \mose{} and \nbse{} orbitals except at the $\Gamma$ point where there is weak mixing. This mixing originates from the fact that Nb and Mo \emph{d}-bands at the $\Gamma$ point have an out of-plane (d$_{z^2}$) orbital character resulting in a weak hybridization, while the high-symmetry $K$ point orbitals are largely in-plane (d$_{xy}$, d$_{x^2 - y^2}$) and therefore less sensitive to the addition of \nbse{}. The lack of Mo-Nb band hybridization at $K$ implies excitonic behaviors in \mose{} will remain in the heterostructure. Interestingly, the main impact of introducing \nbse{} to \mose{} is a downwards shift in the conduction and valence band energies by 0.34 eV, implying that the electron affinity for \mose{} in the heterostructure is $\chi_{hetero}$ = 4.26 eV. This band offset between ML-\mose{} and \nbse{} - \mose{} is a direct consequence of the vdW interaction, where an attractive potential offered by \nbse{} relaxes the energy landscape of the semiconductor (details in Supporting Information - S5). These results are consistent with prior examinations of heterostructure where \nbse{} is stacked upon a semiconductor.\cite{Guan2017b,Lv2018a}. Though insignificant, the observed change in electronic affinity $\chi$ between the two systems will be critical in explaining the origin of L1 peak.

It is important to note here, that while Fermi levels can be tuned arbitrarily within the band-gap of the semiconductor and are pinned by the metal in a metal-semiconductor junction, electron affinities associated with these regions are fixed and determine band alignment. In a scenario shown in Figure \ref{traps}a, the values $\Phi$ and $\chi$ obtained from our first-principles calculations are used to illustrate an in-plane band schematic as one moves from the pristine \mose{} region to the heterostructure. As a consequence of charge rearrangement to electronically equilibriate ML-\mose{} and to match $E_F$ across the two regions, the \mose{} bands in the two regions must bend at the interface. Since $E_g$ does not change, band bending is determined by the differences between $\Phi$ and $\chi$ at the interface. Our DFT calculations indicate that the respective changes in these parameters are different which leads to a band alignment that hosts potential well traps for both electrons and holes as illustrated in Figure \ref{traps}b. This in-plane confinement potential acts as a trap for excitons and trions where we believe they then recombine and result in the L1 peak. A similar mechanism of interface trapping and recombination is known to produce an additional PL feature in AlGaAs/GaAs heterostructures referred to as an 'H-band'.\cite{Yuan1985}

\begin{figure}[h]
    \centering
    \includegraphics[width=3.25in]{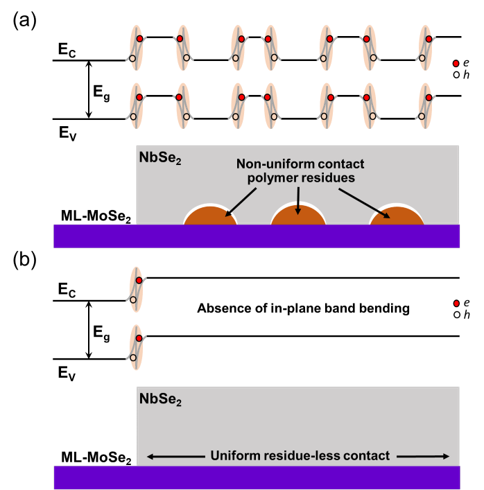}
    \caption{Spatially-resolved band bending schematics. (a) Samples 1 and 2 have large polymer bubbles between \nbse{} and \mose{} that create numerous in-plane boundaries between well-coupled and poorly-coupled regions. PL from L1 is therefore expected from traps randomly distributed across the \nbse{} - \mose{} overlap. (b) Samples 3 and 4 were subjected to an additional cleaning step, 'nano-squeegeeing', that provides a relatively polymer-free interface. L1 therefore only appears at the \nbse{} - \mose{} perimeter.}
    \label{trapdiagram}
\end{figure}

The trapping potential created by band bending provides a natural explanation for the hyperspectral PL maps presented in Figures \ref{linecuts}c and \ref{linecuts}d which correspond to Sample 2 (bare) and Sample 4 (hBN-encapsulated), respectively. Sample 2 was subjected to a vacuum annealing procedure post-assembly that consolidates residual polymer in a vdW heterostructure to form larger bubbles while providing some regions with pristine contact.\cite{Khestanova2016,Frisenda2018a,Onodera2019,Tyurnina2019} This results in a 'patchy' vertical interface where the interlayer separation varies across the \nbse{} - \mose{} heterostructure. Traps will appear at any boundary between well-coupled and poorly-coupled regions that in turn leads to L1 PL emission that varies over the \nbse{} - \mose{} contact (Figure \ref{linecuts}c). In contrast, Sample 4 is fully-encapsulated in hBN and was subjected to the 'nano-squeegee' cleaning technique where an AFM tip is used to physically remove polymer residue from the heterostructure.\cite{Rosenberger2018} This results in a more uniform spacing, and therefore coupling, between \nbse{} and \mose{}. Thus, the only regions that contribute to the band bending required to create a trapping potential are along the perimeter of the \nbse{} - \mose{} interface. The resulting map of the L1 feature therefore should only show PL from the edge of \nbse{} which is exactly the case shown in Figure \ref{linecuts}d. These two situations are schematically illustrated in Figure \ref{trapdiagram}.

\subsection{Localized nature of L1}

\begin{figure}[h]
    \centering
    \includegraphics[width=6.5in]{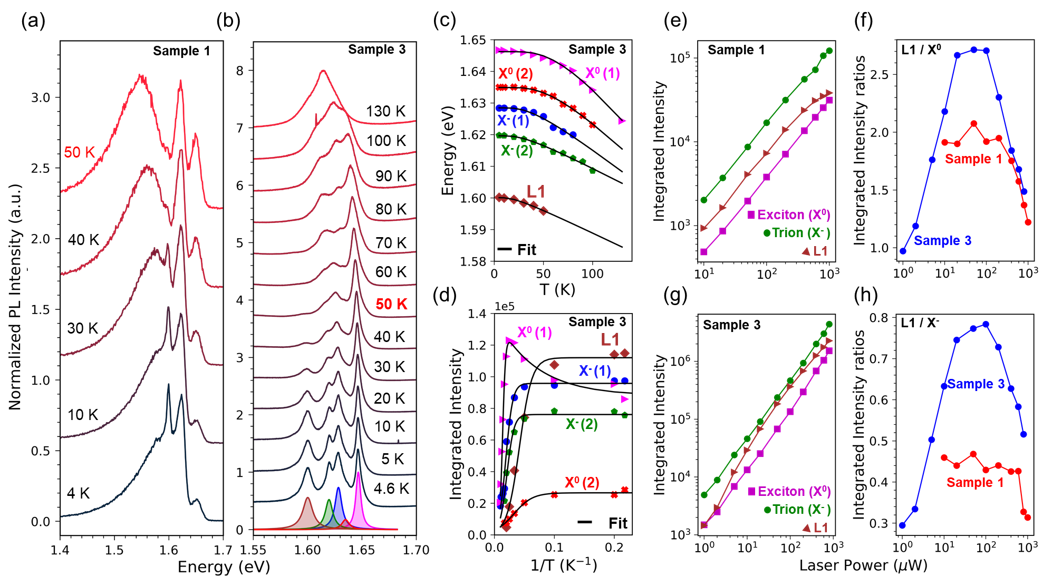}
    \caption{Temperature-dependent PL spectra for (a) Sample 1 and (b) Sample 3 on the \nbse{} - \mose{} interface. Panel (b) also includes an example fit for the spectrum at 4.6 K, identifying five distinct transitions: L1, \xo{}(1), \xo{}(2), \xt{}(1) and \xt{}(2). The duplicate exciton and trion features are believed to originate from intralayer moire\'{e} excitons (Supporting Information). (c) Peak energies extracted from the fits in (b) versus temperature. Marker colors match the fill colors for specific lorentzians in (b) and the black line is a fit to Eq.~\eqref{eq:1}. (d) Temperature-dependent integrated PL intensity of the peaks extracted from (b). The black line is a fit to Eq.~\eqref{eq:2}. Power-dependence of the integrated intensity of L1, \xo{} and \xt{} features plotted for (e) Sample 1 and (g) Sample 3. L1 exhibits a sub-linear power dependence in contrast to \xo{} and \xt. This is depicted by the integrated ratios of (f) L1/\xo{} and (h) L1/\xt{} plotted for both samples.}
    \label{temppower}
\end{figure}

The present physical model implies that L1 originates from the recombination of excitons, and possibly trions, trapped in a confinement potential created by band bending between \mose{} and \nbse{} - \mose{} regions. L1 should therefore exhibit the characteristics of a localized exciton in temperature- and power-dependent PL measurements. Prior observations of localized excitons in 2D materials show they disappear quickly with increasing temperature and exhibit a sub-linear power-dependence due to the saturation of all available trap sites.\cite{Barbone2018,Shibata2005} We find that L1 exhibits both of these characteristics as shown in Figure \ref{temppower}. Temperature-dependent PL measurements on the \nbse{} - \mose{} interface for Samples 1 and 3 are shown in Figure \ref{temppower}a and Figure \ref{temppower}b, respectively. We observe that L1 is weakly robust to changes in temperature (in both samples), especially in the low-temperature regime of 4.6 K - 60 K, contrary to the \xo\ and \xt\ features. This is in agreement with the localized exciton picture, where a thermal energy greater than the trapping potential can eject confined charge carriers through non-radiative mechanisms. Sample 3 has been subjected to an extensive cleaning procedure and is encapsulated in hBN, which results in the observation of five peaks in total compared to Sample 1. These new features are suspected to originate from intralayer moir\'{e} excitons\cite{Zhang2018b} and are elaborated on in the Supporting Information. We fit each spectrum (for Sample 3 - panel b) to a combination of lorentzian functions with a constant background from 1.55 - 1.7 eV to extract the emission energy and the integrated intensity of the observed PL features. These are plotted in Figures \ref{temppower}c and \ref{temppower}d as a function of temperature for peaks assigned as L1, \xo{}(1), \xo{}(2), \xt{}(1) and \xt{}(2) in keeping with the moir\'{e} exciton picture.\cite{Zhang2018b} A representative lorentzian fit is shown below the 4.6 K spectrum in Figure \ref{temppower}b, where the fill color of each peak corresponds to the marker colors in Figure \ref{temppower}c and \ref{temppower}d. In Figure \ref{temppower}c, the extracted peak energies are fit to a standard model that describes temperature-dependence of the semiconductor band-gap,\cite{Ross2013,ODonnell1991} 

\begin{equation}
E_g(T) = E_g(0) - S\langle \hbar\omega \rangle\left[\coth\left(\frac{\langle \hbar\omega \rangle}{2k_BT}\right) -1\right], \label{eq:1}
\end{equation}

\noindent{}where $E_g(0)$ is the transition energy at $T$ = 0 K, $k_B$ is the Boltzmann constant, $S$ is a dimensionless constant describing the strength of the electron-phonon coupling and $\langle \hbar\omega \rangle$ represents the average acoustic phonon energy involved in electron-phonon interactions. We are able to fit all five observed peaks to this functional form and summarize the fitting parameters in Table \ref{Varshinidata}. The obtained values for the dominant \xo{}-like and \xt{}-like peaks agree reasonably well with previous study of pristine ML-\mose.\cite{Ross2013} Interestingly, L1 exhibits the lowest values of $S$, indicative of a weakly-bound confinement of charge carriers that contribute towards L1.

Temperature-dependence of the integrated PL intensity provides key insight into the localized nature of L1. In TMDs with inherent defects and vacancies, thermal disassociation of trapped exciton can sometimes elevate the number of charge carriers available for radiative recombination, giving a rise in the PL intensity up to a certain thermal threshold. We find that \xo(1) exhibits such a non-monotonic temperature-dependent intensity at the interface as shown in Figure \ref{temppower}d. For all other peaks, we observe a monotonic behavior that follows the expectation for localized excitons. Temperature-dependence of the integrated PL intensity can be fit with a modified Arrhenius formula,\cite{Shibata1998,Huang2016,Lin2012}

\begin{equation}
I(T) = I(0)\frac{1 + A e^{-E_{a1}/k_BT}}{1 + B e^{-E_{a2}/k_BT}}, \label{eq:2}
\end{equation}

\noindent{}where $I(0)$ is the integrated PL intensity at $T$ = 0 K, $A$ and $B$ are fitting parameters used to determine the ratio of radiative to non-radiative lifetimes of charge carriers. $E_{a1}$ is the activation energy that increases the number of carriers available for recombination and $E_{a2}$ is the activation energy for the normal thermal quenching process at higher temperatures through non-radiative channels. \xo(1) is fit extremely well by this model as shown in Figure \ref{temppower}d. For the remaining peaks, all of which lie at lower emission energy, we obtain good fits to Eq. \eqref{eq:2} when $A=0$ (Figure \ref{temppower}d) implying remaining peaks are not enhanced by the thermal activation of additional carriers. This suggests that it is likely the thermal quenching of L1, \xo(2), \xt{}(1), and \xt{}(2) states that provides the carriers responsible for the temperature dependent intensity of \xo(1). The fitting parameters for the temperature-dependent integrated PL intensity data are summarized in Table \ref{Arrheniusdata}. The values of $E_{a1}$ = 0.13 meV and $E_{a2}$ = 35 meV for \xo(1) obtained from our fits as well as the values of $E_{a2}$ for other \xo\ and \xt\ species are within the ballpark of similar studies conducted on excitonic states in ML TMD semiconductors,\cite{Huang2016,Mouri2017,Chen2018b} whereas for L1, a value of $E_{a2}$ = 8 meV is extracted. This is much lower than the corresponding values for the dominant \xo{}-like and \xt{}-like peaks (Table \ref{Arrheniusdata}), proving L1 behaves like a weakly-bound trapped-state that disassociates much quicker than the other excitonic states. Additionally, the value $B$, which relates to the ratio of non-radiative decay time in the high-temperature limit to the radiative decay time is also higher for L1 compared to \xo(2) and the prominent \xt(1) and \xt(2) peaks. These values also support our hypothesis that with increasing temperature, rate of non-radiative recombination is much higher for L1 compared to the localized \xt\ states as it originates from confined excitons trapped by the potential wells at the in-plane interface. On the other hand, the value of $E_{a2}$ for \xo(1) refers to the activation energy that is needed to thermally quench as the temperature is increased up to 300 K and equals the binding energy required for exciton disassociation. Since in our case, the dominant \xo(1) peak does not fully quench and our temperature range does not go beyond 130 K, the value obtained from our fit is only the lower bound of the exciton binding energy in a \nbse{} - \mose{} heterostructure. We have included similar analysis for a ML-\mose{} sample in the Supporting Information (S6), where we cover a wider temperature range (5 K - 300 K) and expect activation and binding energies for \xt\ to closely match for a typical monolayer.

\begin{table*}
	\caption{Best-fit parameters of Eq. \eqref{eq:1} to the temperature-dependent emission energies of Sample 3 (Figure \ref{temppower}c).}
	\vspace{5pt}
	\label{Varshinidata}
	\begin{tabular}{c c c c}
		\hline
		Peak ID & $E_g$ (eV) & $S$ & $\langle \hbar\omega \rangle$ (eV)\\
		\hline\\
	    \vspace{2pt}
		\xo(1) & 1.646 & 2.43 & 0.018\\
		\vspace{2pt}
		\xo(2) & 1.635 & 1.87 & 0.015\\
		\vspace{2pt}
		\xt(1) & 1.628 & 1.40  & 0.01\\
		\vspace{2pt}
		\xt(2) & 1.619 & 0.94 & 0.007\\
		\vspace{2pt}
		L1 & 1.60 & 0.86 & 0.005\\
		\hline
	\end{tabular}
\end{table*}

\begin{table*}
	\caption{Best-fit parameters of Eq. \eqref{eq:2} to the temperature-dependent integrated PL intensity of Sample 3 (Figure \ref{temppower}d)}
	\vspace{5pt}
	\label{Arrheniusdata}
	\begin{tabular}{c c c c c c }
		\hline
		Peak ID & $E_{a1}$ (meV) &  $E_{a2}$ (meV) & $A$ & $B$ & $I(0)$\\
	    \hline\\
	    \vspace{2pt}
		\xo(1) & 0.12 & 34.43 & 0.58 & 237.62 & 87833.4\\
		\vspace{2pt}
		\xo(2) & - & 5.53 & - & 8.09 & 26604.3\\
		\vspace{2pt}
		\xt(1) & - & 15.00 & - & 26.73 & 95801.4\\
		\vspace{2pt}
		\xt(2) & - & 14.50 & - & 33.06 & 76111.4\\
		\vspace{2pt}
		L1 & - & 8.30 & - & 56.10 & 112037\\
		\hline
	\end{tabular}
\end{table*}

Power-dependent measurements are presented in Figure \ref{temppower}e and \ref{temppower}g for Sample 1 and Sample 3, respectively. Here, the PL integrated intensity of \xo{} and \xt{} scales linearly with increasing power, but L1 exhibits a sub-linear behavior indicating saturation at higher laser powers. This can be made more obvious by examining the power dependence of the integrated intensity ratio of L1 to \xo{} and \xt{} (Figure \ref{temppower}f and \ref{temppower}h), which shows that L1 initially grows more rapidly compared to the excitonic states, but the rate eventually slows down and plateaus at higher excitation power. This is typical of a defect-like trap state, where increased carrier injection saturates the trap site and manifests as a drop in the relative ratio to the excitonic features at increasing laser power. Similar observation are seen for Sample 1, although lack of data at low-excitation power prevents us from a qualitative comparison in the low-power regime.

\section{Conclusion}
In this study we have identified a new spectral feature, L1, in \nbse{} - \mose{} vdW heterostructures. We attribute this feature to recombination of localized excitons trapped by in-plane confinement potentials formed between \nbse{} - \mose{} and \mose{}. Both our first-principles calculations as well the temperature- and power-dependent PL measurements support this interpretation. Surprisingly, even under widely different fabrication procedures of our heterostructures, L1 remains robust and appears consistently at $\approx$ 1.6 eV. We compare our findings to other low-energy emissions observed in \mose{}-based vdW heterostructures, and show that L1 is spectrally unique to the geometry discussed in this manuscript, but may occur in a variety of metal-semiconductor heterostructures depending on the in-plane band offsets between the constituent TMDs. This spectral robustness may solve an outstanding problem of spectral inhomogeneity in 2D single-photon sources, where photon energies vary due to the different strain and defect induced confinement potentials. Since metal-semiconductors interfaces are common building blocks for many optoelectronic devices,\cite{LarryA.ColdrenScottW.Corzine2012,Chuang2009} we also envision extending our findings to systems which would explicitly take into account spin and valley degrees of freedom. For example, preserving similar robust optical properties in ferromagnetic metal-semiconductor heterostructures could be an important step in realizing two-dimensional spin-lasers that in common III-V semiconductors can exceed the performance of the best conventional lasers.\cite{Lindemann2019} Furthermore, with high-quality and tunable interfacial properties, similar heterostructures would be an important platform in designing proximitized materials.\cite{Zutic2019}

\section{\label{methods}Methods}

\subsection{Growth of TMD materials}

\nbse{} and \mose{} single crystals were grown by the chemical vapor transport (CVT) method. Polycrystalline \nbse{} and \mose{} precursors were synthesized by reacting stoichiometric amounts of Nb (Mo) and Se powders in vacuum-sealed quartz ampoule at 850 $\degree$C for 72 h. For \nbse{}, a quart ampoule containing about 1g of \nbse{} charge and $\approx$ 80 mg (3.7 mg/cm$^3$) of I$_2$ transport agent was sealed under vacuum and placed in a single-zone furnace. Temperatures at the charge and growth zones were T$_{hot}$ = 825 $\degree$C and T$_{cold}$ = 700 $\degree$C, respectively, and the growth duration was 140 h. 2H-\mose{} single crystals were grown using SeBr$_4$ transport agent (5 mg/cm$^3$) at T$_{hot}$ = 980 $\degree$C and T$_{cold}$ = 850 $\degree$C.

\subsection{Sample preparation, assembly and AFM 'Nano-Squeegee'}
MLs of 2H-\mose{} were obtained through mechanical exfoliation of as-grown crystals and subsequently transferred on O$_2$ plasma cleaned Si/SiO$_2$ substrates (Samples 1 and 2) or on hBN (sample 3 and 4) using the PDMS-based dry viscoelastic stamping method.\cite{Castellanos-Gomez2014} This method has been known to leave polymer residue between the interfaces of two TMDs deposited during the transfer process, but has proven to be optimal for fabricating heterostructures.\cite{Frisenda2018a} Similar procedure was employed to exfoliate and transfer bulk \nbse{} flakes to assemble a heterostructure with bulk \nbse{} - \mose{} stacked vertically. For Samples 1 and 2, no intermediate cleaning procedures were employed to get rid of any polymer residue deposited during the heterostructure assembly process. However, they were annealed later 200 $\degree$C for 5 hrs. under high-vacuum condition. In case of Sample 3 and 4, we performed the AFM-based "nano-squeegee" procedure to optimally create clean interfaces.\cite{Rosenberger2018} This involves the use of a standard AFM tip to push out polymer residue deposited between the two TMDs in a vertical heterostructure. We were able to use the same method to also remove surface residue present on the ML \mose{} flake, using a 7 N/m spring constant ACSTGG-10 tip and a contact force of F = 140 nN. We confirm the success of our procedure by performing PL measurements before and after employing this routine (Supporting information S7). Bulk \nbse{} is then brought into contact with sample and immediately capped with a thin layer of hBN to protect from further degradation. Samples 3 and 4 were additionally vacuum annealed at 200 $\degree$C for 5 hrs. under high-vacuum conditions to observe changes in interfacial effects and related PL properties. Sample fabrication details for Samples 5-10 are included in Supporting Information (S1).

\subsection{Experimental setup}
Low-temperature PL and Raman measurements were carried out on a home-built confocal microscope setup with 532 nm (2.4 $\mu$m spot size) laser excitation focused through a 0.42 NA, 50x long working-distance objective. The light is collected in a back-scattering geometry, with the collection fiber-coupled to a 500 mm focal length single spectrometer integrated with a liquid-N$_2$ cooled CCD detector. The samples were placed under vacuum and cooled in a closed cycle He-cooled cryostat (Montana Instruments Corporation) with a variable temperature range from 4 K - 300 K. Raman measurements of \nbse{} included in the Supporting Information (S4) were carried out using the same collection scheme, however, the excitation path included a collection of Bragg grating notch filters to get us within 15 cm$^{-1}$ of the laser line. The excitation wavelength used for Raman measurements was 532 nm and the laser power was kept below 300 $\mu$W pre-objective.

\subsection{DFT calculations}
The electronic structures of monolayers \nbse, \mose{} and the heterostructure were investigated using the general potential linearized augmented plane-wave (LAPW) method\cite{Singh2006} as implemented in the WIEN2K code.\cite{Blaha2001} The convergence of the calculations regarding the size of the basis set is achieved using an RMT * Kmax value of 7, where RMT is the smallest atomic sphere radius in the unit cell and Kmax is the magnitude of the largest K wave vector inside the first Brillouin zone (BZ). The Perdew-Burke-Ernzerhof generalized-gradient approximation (PBE-GGA)\cite{Perdew1996} is used to describe the exchange and correlation functional. A vacuum space larger than 20 \si{\angstrom} is set to avoid the interaction between the two adjacent layers. The Monkhorst-Pack k-grid of 18x18x1 is adopted for the first BZ integral and the convergence criterion for the charge difference was less than 0.0001e per unit cell. To correctly describe the vdW interactions, we employed empirical pairwise corrections proposed by Grimme\cite{Allouche2012} in terms of the DFT+D2 scheme. Structures were fully relaxed until the force on each atom was less than 0.01 eV/\si{\angstrom}. The optimized lattice constants of \mose{} and \nbse{} are 3.33 \si{\angstrom} and 3.48 \si{\angstrom}, in good agreement with the recent work.\cite{Lv2018a} Since there is a lattice mismatch between \mose{} and \nbse{} monolayers, \nbse{} is applied a strain of -4.3 \% in the heterostructure.

\subsection{Acknowledgement}
P.M.V. and J.J. acknowledge support from the National Science Foundation (NSF) under Grant No. DMR-1748650 and DMR-1847782, and the George Mason University Quantum Materials Center. I.Z and T.Z acknowledge support from the U.S. Department of Energy, Office of Science, Basic Energy Sciences under Award No. DE- SC0004890 and the University at Buffalo Center for Computational Research. This work is partly supported through the Material Genome Initiative funding allocated to NIST. The authors would further like to thank I. Mazin and A. Rigosi for valuable discussions.

\subsection{Disclaimer}
Certain commercial equipment, instruments, or materials are identified in this paper in order to specify the experimental procedure adequately. Such identification is not intended to imply recommendation or endorsement by the National Institute of Standards and Technology, nor is it intended to imply that the materials or equipment identified are necessarily the best available for the purpose.

\bibstyle{achemso}

\providecommand{\latin}[1]{#1}
\makeatletter
\providecommand{\doi}
  {\begingroup\let\do\@makeother\dospecials
  \catcode`\{=1 \catcode`\}=2\doi@aux}
\providecommand{\doi@aux}[1]{\endgroup\texttt{#1}}
\makeatother
\providecommand*\mcitethebibliography{\thebibliography}
\csname @ifundefined\endcsname{endmcitethebibliography}
  {\let\endmcitethebibliography\endthebibliography}{}


\begin{mcitethebibliography}{80}
\providecommand*\natexlab[1]{#1}
\providecommand*\mciteSetBstSublistMode[1]{}
\providecommand*\mciteSetBstMaxWidthForm[2]{}
\providecommand*\mciteBstWouldAddEndPuncttrue
  {\def\EndOfBibitem{\unskip.}}
\providecommand*\mciteBstWouldAddEndPunctfalse
  {\let\EndOfBibitem\relax}
\providecommand*\mciteSetBstMidEndSepPunct[3]{}
\providecommand*\mciteSetBstSublistLabelBeginEnd[3]{}
\providecommand*\EndOfBibitem{}
\mciteSetBstSublistMode{f}
\mciteSetBstMaxWidthForm{subitem}{(\alph{mcitesubitemcount})}
\mciteSetBstSublistLabelBeginEnd
  {\mcitemaxwidthsubitemform\space}
  {\relax}
  {\relax}

\bibitem[Alferov(2001)]{Alferov2002}
Alferov,~Z.~I. {Nobel Lecture: The double heterostructure concept and its
  applications in physics, electronics, and technology}. \emph{Rev. Mod. Phys.}
  \textbf{2001}, \emph{73}, 767--782\relax
\mciteBstWouldAddEndPuncttrue
\mciteSetBstMidEndSepPunct{\mcitedefaultmidpunct}
{\mcitedefaultendpunct}{\mcitedefaultseppunct}\relax
\EndOfBibitem
\bibitem[{\v{Z}}uti{\'{c}} \latin{et~al.}(2019){\v{Z}}uti{\'{c}},
  Matos-Abiague, Scharf, Dery, and Belashchenko]{Zutic2019}
{\v{Z}}uti{\'{c}},~I.; Matos-Abiague,~A.; Scharf,~B.; Dery,~H.;
  Belashchenko,~K. {Proximitized materials}. \emph{Mater. Today} \textbf{2019},
  \emph{22}, 85--107\relax
\mciteBstWouldAddEndPuncttrue
\mciteSetBstMidEndSepPunct{\mcitedefaultmidpunct}
{\mcitedefaultendpunct}{\mcitedefaultseppunct}\relax
\EndOfBibitem
\bibitem[Choi \latin{et~al.}(2017)Choi, Choudhary, Han, Park, Akinwande, and
  Lee]{Choi2017}
Choi,~W.; Choudhary,~N.; Han,~G.~H.; Park,~J.; Akinwande,~D.; Lee,~Y.~H.
  {Recent development of two-dimensional transition metal dichalcogenides and
  their applications}. \emph{Mater. Today} \textbf{2017}, \emph{20},
  116--130\relax
\mciteBstWouldAddEndPuncttrue
\mciteSetBstMidEndSepPunct{\mcitedefaultmidpunct}
{\mcitedefaultendpunct}{\mcitedefaultseppunct}\relax
\EndOfBibitem
\bibitem[Manzeli \latin{et~al.}(2017)Manzeli, Ovchinnikov, Pasquier, Yazyev,
  and Kis]{Manzeli2017}
Manzeli,~S.; Ovchinnikov,~D.; Pasquier,~D.; Yazyev,~O.~V.; Kis,~A. {2D
  transition metal dichalcogenides}. \emph{Nat. Rev. Mater.} \textbf{2017},
  \emph{2}, 17033\relax
\mciteBstWouldAddEndPuncttrue
\mciteSetBstMidEndSepPunct{\mcitedefaultmidpunct}
{\mcitedefaultendpunct}{\mcitedefaultseppunct}\relax
\EndOfBibitem
\bibitem[Mak and Shan(2016)Mak, and Shan]{Mak2016}
Mak,~K.~F.; Shan,~J. {Photonics and optoelectronics of 2D semiconductor
  transition metal dichalcogenides}. \emph{Nat. Photonics} \textbf{2016},
  \emph{10}, 216--226\relax
\mciteBstWouldAddEndPuncttrue
\mciteSetBstMidEndSepPunct{\mcitedefaultmidpunct}
{\mcitedefaultendpunct}{\mcitedefaultseppunct}\relax
\EndOfBibitem
\bibitem[Guguchia \latin{et~al.}(2018)Guguchia, Kerelsky, Edelberg, Banerjee,
  von Rohr, Scullion, Augustin, Scully, Rhodes, Shermadini, Luetkens,
  Shengelaya, Baines, Morenzoni, Amato, Hone, Khasanov, Billinge, Santos,
  Pasupathy, and Uemura]{Guguchia2018}
Guguchia,~Z.; Kerelsky,~A.; Edelberg,~D.; Banerjee,~S.; von Rohr,~F.;
  Scullion,~D.; Augustin,~M.; Scully,~M.; Rhodes,~D.~A.; Shermadini,~Z.
  \latin{et~al.}  {Magnetism in semiconducting molybdenum dichalcogenides}.
  \emph{Sci. Adv.} \textbf{2018}, \emph{4}, eaat3672\relax
\mciteBstWouldAddEndPuncttrue
\mciteSetBstMidEndSepPunct{\mcitedefaultmidpunct}
{\mcitedefaultendpunct}{\mcitedefaultseppunct}\relax
\EndOfBibitem
\bibitem[Mak \latin{et~al.}(2010)Mak, Lee, Hone, Shan, and Heinz]{Mak2010}
Mak,~K.~F.; Lee,~C.; Hone,~J.; Shan,~J.; Heinz,~T.~F. {Atomically Thin
  MoS$_{2}$: A New Direct-Gap Semiconductor}. \emph{Phys. Rev. Lett.}
  \textbf{2010}, \emph{105}, 136805\relax
\mciteBstWouldAddEndPuncttrue
\mciteSetBstMidEndSepPunct{\mcitedefaultmidpunct}
{\mcitedefaultendpunct}{\mcitedefaultseppunct}\relax
\EndOfBibitem
\bibitem[Chernikov \latin{et~al.}(2014)Chernikov, Berkelbach, Hill, Rigosi, Li,
  Aslan, Reichman, Hybertsen, and Heinz]{Chernikov2014}
Chernikov,~A.; Berkelbach,~T.~C.; Hill,~H.~M.; Rigosi,~A.; Li,~Y.;
  Aslan,~O.~B.; Reichman,~D.~R.; Hybertsen,~M.~S.; Heinz,~T.~F. {Exciton
  Binding Energy and Nonhydrogenic Rydberg Series in Monolayer WS$_{2}$}.
  \emph{Phys. Rev. Lett.} \textbf{2014}, \emph{113}, 076802\relax
\mciteBstWouldAddEndPuncttrue
\mciteSetBstMidEndSepPunct{\mcitedefaultmidpunct}
{\mcitedefaultendpunct}{\mcitedefaultseppunct}\relax
\EndOfBibitem
\bibitem[Wang \latin{et~al.}(2018)Wang, Chernikov, Glazov, Heinz, Marie, Amand,
  and Urbaszek]{Wang2017a}
Wang,~G.; Chernikov,~A.; Glazov,~M.~M.; Heinz,~T.~F.; Marie,~X.; Amand,~T.;
  Urbaszek,~B. {Colloquium : Excitons in atomically thin transition metal
  dichalcogenides}. \emph{Rev. Mod. Phys.} \textbf{2018}, \emph{90},
  021001\relax
\mciteBstWouldAddEndPuncttrue
\mciteSetBstMidEndSepPunct{\mcitedefaultmidpunct}
{\mcitedefaultendpunct}{\mcitedefaultseppunct}\relax
\EndOfBibitem
\bibitem[Mueller and Malic(2018)Mueller, and Malic]{Mueller2018}
Mueller,~T.; Malic,~E. {Exciton physics and device application of
  two-dimensional transition metal dichalcogenide semiconductors}. \emph{npj 2D
  Mater. Appl.} \textbf{2018}, \emph{2}\relax
\mciteBstWouldAddEndPuncttrue
\mciteSetBstMidEndSepPunct{\mcitedefaultmidpunct}
{\mcitedefaultendpunct}{\mcitedefaultseppunct}\relax
\EndOfBibitem
\bibitem[Xu \latin{et~al.}(2014)Xu, Yao, Xiao, and Heinz]{Xu2014}
Xu,~X.; Yao,~W.; Xiao,~D.; Heinz,~T.~F. {Spin and pseudospins in layered
  transition metal dichalcogenides}. \emph{Nat. Phys.} \textbf{2014},
  \emph{10}, 343--350\relax
\mciteBstWouldAddEndPuncttrue
\mciteSetBstMidEndSepPunct{\mcitedefaultmidpunct}
{\mcitedefaultendpunct}{\mcitedefaultseppunct}\relax
\EndOfBibitem
\bibitem[Jariwala \latin{et~al.}(2017)Jariwala, Marks, and
  Hersam]{Jariwala2017}
Jariwala,~D.; Marks,~T.~J.; Hersam,~M.~C. {Mixed-dimensional van der Waals
  heterostructures}. \emph{Nat. Mater.} \textbf{2017}, \emph{16},
  170--181\relax
\mciteBstWouldAddEndPuncttrue
\mciteSetBstMidEndSepPunct{\mcitedefaultmidpunct}
{\mcitedefaultendpunct}{\mcitedefaultseppunct}\relax
\EndOfBibitem
\bibitem[Liu \latin{et~al.}(2016)Liu, Weiss, Duan, Cheng, Huang, and
  Duan]{Liu2016b}
Liu,~Y.; Weiss,~N.~O.; Duan,~X.; Cheng,~H.-C.; Huang,~Y.; Duan,~X. {Van der
  Waals heterostructures and devices}. \emph{Nat. Rev. Mater.} \textbf{2016},
  \emph{1}, 16042\relax
\mciteBstWouldAddEndPuncttrue
\mciteSetBstMidEndSepPunct{\mcitedefaultmidpunct}
{\mcitedefaultendpunct}{\mcitedefaultseppunct}\relax
\EndOfBibitem
\bibitem[Liu \latin{et~al.}(2012)Liu, Neal, and Ye]{Liu2012}
Liu,~H.; Neal,~A.~T.; Ye,~P.~D. {Channel length scaling of MoS$_{2}$ MOSFETs}.
  \emph{ACS Nano} \textbf{2012}, \emph{6}, 8563--8569\relax
\mciteBstWouldAddEndPuncttrue
\mciteSetBstMidEndSepPunct{\mcitedefaultmidpunct}
{\mcitedefaultendpunct}{\mcitedefaultseppunct}\relax
\EndOfBibitem
\bibitem[Das and Appenzeller(2013)Das, and Appenzeller]{Das2013a}
Das,~S.; Appenzeller,~J. {WSe$_{2}$ field effect transistors with enhanced
  ambipolar characteristics}. \emph{Appl. Phys. Lett.} \textbf{2013},
  \emph{103}, 103501\relax
\mciteBstWouldAddEndPuncttrue
\mciteSetBstMidEndSepPunct{\mcitedefaultmidpunct}
{\mcitedefaultendpunct}{\mcitedefaultseppunct}\relax
\EndOfBibitem
\bibitem[Larentis \latin{et~al.}(2012)Larentis, Fallahazad, and
  Tutuc]{Larentis2012}
Larentis,~S.; Fallahazad,~B.; Tutuc,~E. {Field-effect transistors and intrinsic
  mobility in ultra-thin MoSe$_{2}$ layers}. \emph{Appl. Phys. Lett.}
  \textbf{2012}, \emph{101}, 223104\relax
\mciteBstWouldAddEndPuncttrue
\mciteSetBstMidEndSepPunct{\mcitedefaultmidpunct}
{\mcitedefaultendpunct}{\mcitedefaultseppunct}\relax
\EndOfBibitem
\bibitem[Das \latin{et~al.}(2013)Das, Chen, Penumatcha, and
  Appenzeller]{Das2013}
Das,~S.; Chen,~H.~Y.; Penumatcha,~A.~V.; Appenzeller,~J. {High performance
  multilayer MoS$_{2}$ transistors with scandium contacts}. \emph{Nano Lett.}
  \textbf{2013}, \emph{13}, 100--105\relax
\mciteBstWouldAddEndPuncttrue
\mciteSetBstMidEndSepPunct{\mcitedefaultmidpunct}
{\mcitedefaultendpunct}{\mcitedefaultseppunct}\relax
\EndOfBibitem
\bibitem[Bhanu \latin{et~al.}(2015)Bhanu, Islam, Tetard, and
  Khondaker]{Bhanu2014}
Bhanu,~U.; Islam,~M.~R.; Tetard,~L.; Khondaker,~S.~I. {Photoluminescence
  quenching in gold - MoS$_{2}$ hybrid nanoflakes}. \emph{Sci. Rep.}
  \textbf{2015}, \emph{4}, 5575\relax
\mciteBstWouldAddEndPuncttrue
\mciteSetBstMidEndSepPunct{\mcitedefaultmidpunct}
{\mcitedefaultendpunct}{\mcitedefaultseppunct}\relax
\EndOfBibitem
\bibitem[Froehlicher \latin{et~al.}(2018)Froehlicher, Lorchat, and
  Berciaud]{Froehlicher2018}
Froehlicher,~G.; Lorchat,~E.; Berciaud,~S. {Charge Versus Energy Transfer in
  Atomically Thin Graphene-Transition Metal Dichalcogenide van der Waals
  Heterostructures}. \emph{Phys. Rev. X} \textbf{2018}, \emph{8}, 011007\relax
\mciteBstWouldAddEndPuncttrue
\mciteSetBstMidEndSepPunct{\mcitedefaultmidpunct}
{\mcitedefaultendpunct}{\mcitedefaultseppunct}\relax
\EndOfBibitem
\bibitem[Zhang \latin{et~al.}(2019)Zhang, Yan, Sun, Dong, Yildirim, Wang, Wen,
  Neupane, Sharma, Zhu, Zhang, Liang, Liu, Nguyen, Macdonald, and
  Lu]{Zhang2019}
Zhang,~L.; Yan,~H.; Sun,~X.; Dong,~M.; Yildirim,~T.; Wang,~B.; Wen,~B.;
  Neupane,~G.~P.; Sharma,~A.; Zhu,~Y. \latin{et~al.}  {Modulated interlayer
  charge transfer dynamics in a monolayer TMD/metal junction}. \emph{Nanoscale}
  \textbf{2019}, \emph{11}, 418--425\relax
\mciteBstWouldAddEndPuncttrue
\mciteSetBstMidEndSepPunct{\mcitedefaultmidpunct}
{\mcitedefaultendpunct}{\mcitedefaultseppunct}\relax
\EndOfBibitem
\bibitem[Guan \latin{et~al.}(2017)Guan, Chuang, Zhou, and
  Tom{\'{a}}nek]{Guan2017b}
Guan,~J.; Chuang,~H.~J.; Zhou,~Z.; Tom{\'{a}}nek,~D. {Optimizing Charge
  Injection across Transition Metal Dichalcogenide Heterojunctions: Theory and
  Experiment}. \emph{ACS Nano} \textbf{2017}, \emph{11}, 3904--3910\relax
\mciteBstWouldAddEndPuncttrue
\mciteSetBstMidEndSepPunct{\mcitedefaultmidpunct}
{\mcitedefaultendpunct}{\mcitedefaultseppunct}\relax
\EndOfBibitem
\bibitem[Shin \latin{et~al.}(2018)Shin, Yoon, Kim, Kim, Lim, Yu, Park, Yi, Kim,
  Jun, and Im]{Shin2018}
Shin,~H.~G.; Yoon,~H.~S.; Kim,~J.~S.; Kim,~M.; Lim,~J.~Y.; Yu,~S.; Park,~J.~H.;
  Yi,~Y.; Kim,~T.; Jun,~S.~C. \latin{et~al.}  {Vertical and In-Plane Current
  Devices Using NbS$_{2}$/n-MoS$_{2}$ van der Waals Schottky Junction and
  Graphene Contact}. \emph{Nano Lett.} \textbf{2018}, \emph{18},
  1937--1945\relax
\mciteBstWouldAddEndPuncttrue
\mciteSetBstMidEndSepPunct{\mcitedefaultmidpunct}
{\mcitedefaultendpunct}{\mcitedefaultseppunct}\relax
\EndOfBibitem
\bibitem[Dvir \latin{et~al.}(2018)Dvir, Massee, Attias, Khodas, Aprili, Quay,
  and Steinberg]{Dvir2018}
Dvir,~T.; Massee,~F.; Attias,~L.; Khodas,~M.; Aprili,~M.; Quay,~C. H.~L.;
  Steinberg,~H. {Spectroscopy of bulk and few-layer superconducting NbSe$_{2}$
  with van der Waals tunnel junctions}. \emph{Nat. Commun.} \textbf{2018},
  \emph{9}, 598\relax
\mciteBstWouldAddEndPuncttrue
\mciteSetBstMidEndSepPunct{\mcitedefaultmidpunct}
{\mcitedefaultendpunct}{\mcitedefaultseppunct}\relax
\EndOfBibitem
\bibitem[Huang \latin{et~al.}(2018)Huang, Narayan, Zhang, Liu, Yan, Wang,
  Zhang, Wang, Zhou, Yi, Liu, Ling, Zhang, Liu, Sankar, Chou, Wang, Shi, Law,
  Sanvito, Zhou, Han, and Xiu]{Huang2018}
Huang,~C.; Narayan,~A.; Zhang,~E.; Liu,~Y.; Yan,~X.; Wang,~J.; Zhang,~C.;
  Wang,~W.; Zhou,~T.; Yi,~C. \latin{et~al.}  {Inducing Strong Superconductivity
  in WTe$_{2}$ by a Proximity Effect}. \emph{ACS Nano} \textbf{2018},
  \emph{12}, 7185--7196\relax
\mciteBstWouldAddEndPuncttrue
\mciteSetBstMidEndSepPunct{\mcitedefaultmidpunct}
{\mcitedefaultendpunct}{\mcitedefaultseppunct}\relax
\EndOfBibitem
\bibitem[Lv \latin{et~al.}(2018)Lv, Wei, Zhao, Li, Huang, and Dai]{Lv2018a}
Lv,~X.; Wei,~W.; Zhao,~P.; Li,~J.; Huang,~B.; Dai,~Y. {Tunable Schottky
  contacts in MSe$_{2}$/NbSe$_{2}$ (M = Mo and W) heterostructures and
  promising application potential in field-effect transistors}. \emph{Phys.
  Chem. Chem. Phys.} \textbf{2018}, \emph{20}, 1897--1903\relax
\mciteBstWouldAddEndPuncttrue
\mciteSetBstMidEndSepPunct{\mcitedefaultmidpunct}
{\mcitedefaultendpunct}{\mcitedefaultseppunct}\relax
\EndOfBibitem
\bibitem[Yabuki \latin{et~al.}(2016)Yabuki, Moriya, Arai, Sata, Morikawa,
  Masubuchi, and Machida]{Yabuki2016}
Yabuki,~N.; Moriya,~R.; Arai,~M.; Sata,~Y.; Morikawa,~S.; Masubuchi,~S.;
  Machida,~T. {Supercurrent in van der Waals Josephson junction}. \emph{Nat.
  Commun.} \textbf{2016}, \emph{7}, 10616\relax
\mciteBstWouldAddEndPuncttrue
\mciteSetBstMidEndSepPunct{\mcitedefaultmidpunct}
{\mcitedefaultendpunct}{\mcitedefaultseppunct}\relax
\EndOfBibitem
\bibitem[Lee \latin{et~al.}(2019)Lee, Kim, Watanabe, Taniguchi, Lee, and
  Lee]{Lee2019}
Lee,~J.; Kim,~M.; Watanabe,~K.; Taniguchi,~T.; Lee,~G.~H.; Lee,~H.~J. {Planar
  graphene Josephson coupling via van der Waals superconducting contacts}.
  \emph{Curr. Appl. Phys.} \textbf{2019}, \emph{19}, 251--255\relax
\mciteBstWouldAddEndPuncttrue
\mciteSetBstMidEndSepPunct{\mcitedefaultmidpunct}
{\mcitedefaultendpunct}{\mcitedefaultseppunct}\relax
\EndOfBibitem
\bibitem[Kim \latin{et~al.}(2017)Kim, Park, Lee, Lee, Park, Lee, Lee, and
  Lee]{Kim2017d}
Kim,~M.; Park,~G.~H.; Lee,~J.; Lee,~J.~H.; Park,~J.; Lee,~H.; Lee,~G.~H.;
  Lee,~H.~J. {Strong Proximity Josephson Coupling in Vertically Stacked
  NbSe$_{2}$-Graphene-NbSe$_{2}$ van der Waals Junctions}. \emph{Nano Lett.}
  \textbf{2017}, \emph{17}, 6125--6130\relax
\mciteBstWouldAddEndPuncttrue
\mciteSetBstMidEndSepPunct{\mcitedefaultmidpunct}
{\mcitedefaultendpunct}{\mcitedefaultseppunct}\relax
\EndOfBibitem
\bibitem[Pan \latin{et~al.}(2016)Pan, Li, Ye, Quhe, Song, Wang, Zheng, Pan,
  Guo, Yang, and Lu]{Pan2016}
Pan,~Y.; Li,~S.; Ye,~M.; Quhe,~R.; Song,~Z.; Wang,~Y.; Zheng,~J.; Pan,~F.;
  Guo,~W.; Yang,~J. \latin{et~al.}  {Interfacial Properties of Monolayer
  MoSe$_{2}$-Metal Contacts}. \emph{J. Phys. Chem. C} \textbf{2016},
  \emph{120}, 13063--13070\relax
\mciteBstWouldAddEndPuncttrue
\mciteSetBstMidEndSepPunct{\mcitedefaultmidpunct}
{\mcitedefaultendpunct}{\mcitedefaultseppunct}\relax
\EndOfBibitem
\bibitem[Liu \latin{et~al.}(2018)Liu, Li, Shi, Zhang, Pan, Ye, Quhe, Wang,
  Zhang, Yan, Xu, Guo, Pan, and Lu]{Liu2018b}
Liu,~S.; Li,~J.; Shi,~B.; Zhang,~X.; Pan,~Y.; Ye,~M.; Quhe,~R.; Wang,~Y.;
  Zhang,~H.; Yan,~J. \latin{et~al.}  {Gate-tunable interfacial properties of
  in-plane ML MX$_{2}$ 1T′--2H heterojunctions}. \emph{J. Mater. Chem. C}
  \textbf{2018}, \emph{6}, 5651--5661\relax
\mciteBstWouldAddEndPuncttrue
\mciteSetBstMidEndSepPunct{\mcitedefaultmidpunct}
{\mcitedefaultendpunct}{\mcitedefaultseppunct}\relax
\EndOfBibitem
\bibitem[{\c{C}}akir \latin{et~al.}(2014){\c{C}}akir, Sevik, and
  Peeters]{Cakir2014}
{\c{C}}akir,~D.; Sevik,~C.; Peeters,~F.~M. {Engineering electronic properties
  of metal-MoSe$_{2}$ interfaces using self-assembled monolayers}. \emph{J.
  Mater. Chem. C} \textbf{2014}, \emph{2}, 9842--9849\relax
\mciteBstWouldAddEndPuncttrue
\mciteSetBstMidEndSepPunct{\mcitedefaultmidpunct}
{\mcitedefaultendpunct}{\mcitedefaultseppunct}\relax
\EndOfBibitem
\bibitem[Huang \latin{et~al.}(2017)Huang, Li, Zhong, Wei, and Li]{Huang2017a}
Huang,~L.; Li,~B.; Zhong,~M.; Wei,~Z.; Li,~J. {Tunable Schottky Barrier at
  MoSe$_{2}$/Metal Interfaces with a Buffer Layer}. \emph{J. Phys. Chem. C}
  \textbf{2017}, \emph{121}, 9305--9311\relax
\mciteBstWouldAddEndPuncttrue
\mciteSetBstMidEndSepPunct{\mcitedefaultmidpunct}
{\mcitedefaultendpunct}{\mcitedefaultseppunct}\relax
\EndOfBibitem
\bibitem[Liu \latin{et~al.}(2016)Liu, Stradins, and Wei]{Liu2016c}
Liu,~Y.; Stradins,~P.; Wei,~S.-H. {Van der Waals metal-semiconductor junction:
  Weak Fermi level pinning enables effective tuning of Schottky barrier}.
  \emph{Sci. Adv.} \textbf{2016}, \emph{2}, e1600069\relax
\mciteBstWouldAddEndPuncttrue
\mciteSetBstMidEndSepPunct{\mcitedefaultmidpunct}
{\mcitedefaultendpunct}{\mcitedefaultseppunct}\relax
\EndOfBibitem
\bibitem[Mönch(1999)]{Monch1999}
Mönch,~W. {Barrier heights of real Schottky contacts explained by
  metal-induced gap states and lateral inhomogeneities}. \emph{J. Vac. Sci.
  Technol. B Microelectron. Nanom. Struct.} \textbf{1999}, \emph{17},
  1867\relax
\mciteBstWouldAddEndPuncttrue
\mciteSetBstMidEndSepPunct{\mcitedefaultmidpunct}
{\mcitedefaultendpunct}{\mcitedefaultseppunct}\relax
\EndOfBibitem
\bibitem[Heine(1965)]{Heine1965}
Heine,~V. {Theory of Surface States}. \emph{Phys. Rev.} \textbf{1965},
  \emph{138}, A1689--A1696\relax
\mciteBstWouldAddEndPuncttrue
\mciteSetBstMidEndSepPunct{\mcitedefaultmidpunct}
{\mcitedefaultendpunct}{\mcitedefaultseppunct}\relax
\EndOfBibitem
\bibitem[Sajjad \latin{et~al.}(2019)Sajjad, Yang, Altermatt, Singh,
  Schwingenschl{\"{o}}gl, and {De Wolf}]{Sajjad2019}
Sajjad,~M.; Yang,~X.; Altermatt,~P.; Singh,~N.; Schwingenschl{\"{o}}gl,~U.; {De
  Wolf},~S. {Metal-induced gap states in passivating metal/silicon contacts}.
  \emph{Appl. Phys. Lett.} \textbf{2019}, \emph{114}, 071601\relax
\mciteBstWouldAddEndPuncttrue
\mciteSetBstMidEndSepPunct{\mcitedefaultmidpunct}
{\mcitedefaultendpunct}{\mcitedefaultseppunct}\relax
\EndOfBibitem
\bibitem[Lu \latin{et~al.}(2014)Lu, Li, Mao, Wang, and Andrei]{Lu2014}
Lu,~C.~P.; Li,~G.; Mao,~J.; Wang,~L.~M.; Andrei,~E.~Y. {Bandgap, mid-gap
  states, and gating effects in MoS$_{2}$}. \emph{Nano Lett.} \textbf{2014},
  \emph{14}, 4628--4633\relax
\mciteBstWouldAddEndPuncttrue
\mciteSetBstMidEndSepPunct{\mcitedefaultmidpunct}
{\mcitedefaultendpunct}{\mcitedefaultseppunct}\relax
\EndOfBibitem
\bibitem[Scharf \latin{et~al.}(2019)Scharf, {Van Tuan}, {\v{Z}}uti{\'{c}}, and
  Dery]{Scharf2019}
Scharf,~B.; {Van Tuan},~D.; {\v{Z}}uti{\'{c}},~I.; Dery,~H. {Dynamical
  screening in monolayer transition-metal dichalcogenides and its
  manifestations in the exciton spectrum}. \emph{J. Phys. Condens. Matter}
  \textbf{2019}, \emph{31}, 203001\relax
\mciteBstWouldAddEndPuncttrue
\mciteSetBstMidEndSepPunct{\mcitedefaultmidpunct}
{\mcitedefaultendpunct}{\mcitedefaultseppunct}\relax
\EndOfBibitem
\bibitem[Ross \latin{et~al.}(2013)Ross, Wu, Yu, Ghimire, Jones, Aivazian, Yan,
  Mandrus, Xiao, Yao, and Xu]{Ross2013}
Ross,~J.~S.; Wu,~S.; Yu,~H.; Ghimire,~N.~J.; Jones,~A.~M.; Aivazian,~G.;
  Yan,~J.; Mandrus,~D.~G.; Xiao,~D.; Yao,~W. \latin{et~al.}  {Electrical
  control of neutral and charged excitons in a monolayer semiconductor}.
  \emph{Nat. Commun.} \textbf{2013}, \emph{4}, 1474\relax
\mciteBstWouldAddEndPuncttrue
\mciteSetBstMidEndSepPunct{\mcitedefaultmidpunct}
{\mcitedefaultendpunct}{\mcitedefaultseppunct}\relax
\EndOfBibitem
\bibitem[Edelberg \latin{et~al.}(2019)Edelberg, Rhodes, Kerelsky, Kim, Wang,
  Zangiabadi, Kim, Abhinandan, Ardelean, Scully, Scullion, Embon, Zu, Santos,
  Balicas, Marianetti, Barmak, Zhu, Hone, and Pasupathy]{Edelberg2019}
Edelberg,~D.; Rhodes,~D.; Kerelsky,~A.; Kim,~B.; Wang,~J.; Zangiabadi,~A.;
  Kim,~C.; Abhinandan,~A.; Ardelean,~J.; Scully,~M. \latin{et~al.}
  {Approaching the Intrinsic Limit in Transition Metal Diselenides via Point
  Defect Control}. \emph{Nano Lett.} \textbf{2019}, \emph{19}, 4371--4379\relax
\mciteBstWouldAddEndPuncttrue
\mciteSetBstMidEndSepPunct{\mcitedefaultmidpunct}
{\mcitedefaultendpunct}{\mcitedefaultseppunct}\relax
\EndOfBibitem
\bibitem[Lu \latin{et~al.}(2019)Lu, Rhodes, Li, Tuan, Jiang, Ludwig, Jiang,
  Lian, Shi, Hone, Dery, and Smirnov]{Lu_2019}
Lu,~Z.; Rhodes,~D.; Li,~Z.; Tuan,~D.~V.; Jiang,~Y.; Ludwig,~J.; Jiang,~Z.;
  Lian,~Z.; Shi,~S.-F.; Hone,~J. \latin{et~al.}  {Magnetic field mixing and
  splitting of bright and dark excitons in monolayer MoSe$_{2}$}. \emph{2D
  Mater.} \textbf{2019}, \emph{7}, 15017\relax
\mciteBstWouldAddEndPuncttrue
\mciteSetBstMidEndSepPunct{\mcitedefaultmidpunct}
{\mcitedefaultendpunct}{\mcitedefaultseppunct}\relax
\EndOfBibitem
\bibitem[Ajayi \latin{et~al.}(2017)Ajayi, Ardelean, Shepard, Wang, Antony,
  Taniguchi, Watanabe, Heinz, Strauf, Zhu, and Hone]{Ajayi2017}
Ajayi,~O.~A.; Ardelean,~J.~V.; Shepard,~G.~D.; Wang,~J.; Antony,~A.;
  Taniguchi,~T.; Watanabe,~K.; Heinz,~T.~F.; Strauf,~S.; Zhu,~X.-Y.
  \latin{et~al.}  {Approaching the intrinsic photoluminescence linewidth in
  transition metal dichalcogenide monolayers}. \emph{2D Mater.} \textbf{2017},
  \emph{4}, 031011\relax
\mciteBstWouldAddEndPuncttrue
\mciteSetBstMidEndSepPunct{\mcitedefaultmidpunct}
{\mcitedefaultendpunct}{\mcitedefaultseppunct}\relax
\EndOfBibitem
\bibitem[Wierzbowski \latin{et~al.}(2017)Wierzbowski, Klein, Sigger,
  Straubinger, Kremser, Taniguchi, Watanabe, Wurstbauer, Holleitner, Kaniber,
  M{\"{u}}ller, and Finley]{Wierzbowski2017}
Wierzbowski,~J.; Klein,~J.; Sigger,~F.; Straubinger,~C.; Kremser,~M.;
  Taniguchi,~T.; Watanabe,~K.; Wurstbauer,~U.; Holleitner,~A.~W.; Kaniber,~M.
  \latin{et~al.}  {Direct exciton emission from atomically thin transition
  metal dichalcogenide heterostructures near the lifetime limit}. \emph{Sci.
  Rep.} \textbf{2017}, \emph{7}, 12383\relax
\mciteBstWouldAddEndPuncttrue
\mciteSetBstMidEndSepPunct{\mcitedefaultmidpunct}
{\mcitedefaultendpunct}{\mcitedefaultseppunct}\relax
\EndOfBibitem
\bibitem[Jadczak \latin{et~al.}(2017)Jadczak, Kutrowska-Girzycka,
  Kapu{\'{s}}ci{\'{n}}ski, Huang, W{\'{o}}js, and Bryja]{Jadczak2017a}
Jadczak,~J.; Kutrowska-Girzycka,~J.; Kapu{\'{s}}ci{\'{n}}ski,~P.; Huang,~Y.~S.;
  W{\'{o}}js,~A.; Bryja,~L. {Probing of free and localized excitons and trions
  in atomically thin WSe$_{2}$, WS$_{2}$, MoSe$_{2}$ and MoS$_{2}$ in
  photoluminescence and reflectivity experiments}. \emph{Nanotechnology}
  \textbf{2017}, \emph{28}, 395702\relax
\mciteBstWouldAddEndPuncttrue
\mciteSetBstMidEndSepPunct{\mcitedefaultmidpunct}
{\mcitedefaultendpunct}{\mcitedefaultseppunct}\relax
\EndOfBibitem
\bibitem[{Van Tuan} \latin{et~al.}(2018){Van Tuan}, Yang, and
  Dery]{VanTuan2018a}
{Van Tuan},~D.; Yang,~M.; Dery,~H. {Coulomb interaction in monolayer
  transition-metal dichalcogenides}. \emph{Phys. Rev. B} \textbf{2018},
  \emph{98}, 125308\relax
\mciteBstWouldAddEndPuncttrue
\mciteSetBstMidEndSepPunct{\mcitedefaultmidpunct}
{\mcitedefaultendpunct}{\mcitedefaultseppunct}\relax
\EndOfBibitem
\bibitem[Florian \latin{et~al.}(2018)Florian, Hartmann, Steinhoff, Klein,
  Holleitner, Finley, Wehling, Kaniber, and Gies]{Florian2018}
Florian,~M.; Hartmann,~M.; Steinhoff,~A.; Klein,~J.; Holleitner,~A.~W.;
  Finley,~J.~J.; Wehling,~T.~O.; Kaniber,~M.; Gies,~C. {The Dielectric Impact
  of Layer Distances on Exciton and Trion Binding Energies in van der Waals
  Heterostructures}. \emph{Nano Lett.} \textbf{2018}, \emph{18},
  2725--2732\relax
\mciteBstWouldAddEndPuncttrue
\mciteSetBstMidEndSepPunct{\mcitedefaultmidpunct}
{\mcitedefaultendpunct}{\mcitedefaultseppunct}\relax
\EndOfBibitem
\bibitem[Zhang \latin{et~al.}(2018)Zhang, Surrente, Baranowski, Maude, Gant,
  Castellanos-Gomez, and Plochocka]{Zhang2018b}
Zhang,~N.; Surrente,~A.; Baranowski,~M.; Maude,~D.~K.; Gant,~P.;
  Castellanos-Gomez,~A.; Plochocka,~P. {Moir{\'{e}} Intralayer Excitons in a
  MoSe$_{2}$/MoS$_{2}$ Heterostructure}. \emph{Nano Lett.} \textbf{2018},
  \emph{18}, 7651--7657\relax
\mciteBstWouldAddEndPuncttrue
\mciteSetBstMidEndSepPunct{\mcitedefaultmidpunct}
{\mcitedefaultendpunct}{\mcitedefaultseppunct}\relax
\EndOfBibitem
\bibitem[Lien \latin{et~al.}(2015)Lien, Kang, Amani, Chen, Tosun, Wang, Roy,
  Eggleston, Wu, Dubey, Lee, He, and Javey]{Lien2015a}
Lien,~D.-H.; Kang,~J.~S.; Amani,~M.; Chen,~K.; Tosun,~M.; Wang,~H.-P.; Roy,~T.;
  Eggleston,~M.~S.; Wu,~M.~C.; Dubey,~M. \latin{et~al.}  {Engineering Light
  Outcoupling in 2D Materials}. \emph{Nano Lett.} \textbf{2015}, \emph{15},
  1356--1361\relax
\mciteBstWouldAddEndPuncttrue
\mciteSetBstMidEndSepPunct{\mcitedefaultmidpunct}
{\mcitedefaultendpunct}{\mcitedefaultseppunct}\relax
\EndOfBibitem
\bibitem[Khestanova \latin{et~al.}(2016)Khestanova, Guinea, Fumagalli, Geim,
  and Grigorieva]{Khestanova2016}
Khestanova,~E.; Guinea,~F.; Fumagalli,~L.; Geim,~A.~K.; Grigorieva,~I.~V.
  {Universal shape and pressure inside bubbles appearing in van der Waals
  heterostructures}. \emph{Nat. Commun.} \textbf{2016}, \emph{7}, 12587\relax
\mciteBstWouldAddEndPuncttrue
\mciteSetBstMidEndSepPunct{\mcitedefaultmidpunct}
{\mcitedefaultendpunct}{\mcitedefaultseppunct}\relax
\EndOfBibitem
\bibitem[Frisenda \latin{et~al.}(2018)Frisenda, Navarro-Moratalla, Gant,
  {P{\'{e}}rez De Lara}, Jarillo-Herrero, Gorbachev, and
  Castellanos-Gomez]{Frisenda2018a}
Frisenda,~R.; Navarro-Moratalla,~E.; Gant,~P.; {P{\'{e}}rez De Lara},~D.;
  Jarillo-Herrero,~P.; Gorbachev,~R.~V.; Castellanos-Gomez,~A. {Recent progress
  in the assembly of nanodevices and van der Waals heterostructures by
  deterministic placement of 2D materials}. \emph{Chem. Soc. Rev.}
  \textbf{2018}, \emph{47}, 53--68\relax
\mciteBstWouldAddEndPuncttrue
\mciteSetBstMidEndSepPunct{\mcitedefaultmidpunct}
{\mcitedefaultendpunct}{\mcitedefaultseppunct}\relax
\EndOfBibitem
\bibitem[Onodera \latin{et~al.}(2019)Onodera, Masubuchi, Moriya, and
  Machida]{Onodera2019}
Onodera,~M.; Masubuchi,~S.; Moriya,~R.; Machida,~T. {Assembly of van der Waals
  heterostructures: exfoliation, searching, and stacking of 2D materials}.
  \emph{Jpn. J. Appl. Phys.} \textbf{2019}, \emph{59}\relax
\mciteBstWouldAddEndPuncttrue
\mciteSetBstMidEndSepPunct{\mcitedefaultmidpunct}
{\mcitedefaultendpunct}{\mcitedefaultseppunct}\relax
\EndOfBibitem
\bibitem[Tyurnina \latin{et~al.}(2019)Tyurnina, Bandurin, Khestanova, Kravets,
  Koperski, Guinea, Grigorenko, Geim, and Grigorieva]{Tyurnina2019}
Tyurnina,~A.~V.; Bandurin,~D.~A.; Khestanova,~E.; Kravets,~V.~G.; Koperski,~M.;
  Guinea,~F.; Grigorenko,~A.~N.; Geim,~A.~K.; Grigorieva,~I.~V. {Strained
  Bubbles in van der Waals Heterostructures as Local Emitters of
  Photoluminescence with Adjustable Wavelength}. \emph{ACS Photonics}
  \textbf{2019}, \emph{6}, 516--524\relax
\mciteBstWouldAddEndPuncttrue
\mciteSetBstMidEndSepPunct{\mcitedefaultmidpunct}
{\mcitedefaultendpunct}{\mcitedefaultseppunct}\relax
\EndOfBibitem
\bibitem[Branny \latin{et~al.}(2016)Branny, Wang, Kumar, Robert, Lassagne,
  Marie, Gerardot, and Urbaszek]{Branny2016}
Branny,~A.; Wang,~G.; Kumar,~S.; Robert,~C.; Lassagne,~B.; Marie,~X.;
  Gerardot,~B.~D.; Urbaszek,~B. {Discrete quantum dot like emitters in
  monolayer MoSe$_{2}$ : Spatial mapping, magneto-optics, and charge tuning}.
  \emph{Appl. Phys. Lett.} \textbf{2016}, \emph{108}, 142101\relax
\mciteBstWouldAddEndPuncttrue
\mciteSetBstMidEndSepPunct{\mcitedefaultmidpunct}
{\mcitedefaultendpunct}{\mcitedefaultseppunct}\relax
\EndOfBibitem
\bibitem[Deilmann and Thygesen(2017)Deilmann, and Thygesen]{Deilmann2017}
Deilmann,~T.; Thygesen,~K.~S. {Dark excitations in monolayer transition metal
  dichalcogenides}. \emph{Phys. Rev. B} \textbf{2017}, \emph{96}, 201113\relax
\mciteBstWouldAddEndPuncttrue
\mciteSetBstMidEndSepPunct{\mcitedefaultmidpunct}
{\mcitedefaultendpunct}{\mcitedefaultseppunct}\relax
\EndOfBibitem
\bibitem[Aas and Bulutay(2018)Aas, and Bulutay]{Aas2018}
Aas,~S.; Bulutay,~C. {Strain dependence of photoluminescence and circular
  dichroism in transition metal dichalcogenides: a k {\textperiodcentered} p
  analysis}. \emph{Opt. Express} \textbf{2018}, \emph{26}, 28672\relax
\mciteBstWouldAddEndPuncttrue
\mciteSetBstMidEndSepPunct{\mcitedefaultmidpunct}
{\mcitedefaultendpunct}{\mcitedefaultseppunct}\relax
\EndOfBibitem
\bibitem[Hao \latin{et~al.}(2017)Hao, Specht, Nagler, Xu, Tran, Singh, Dass,
  Sch{\"{u}}ller, Korn, Richter, Knorr, Li, and Moody]{Hao2017a}
Hao,~K.; Specht,~J.~F.; Nagler,~P.; Xu,~L.; Tran,~K.; Singh,~A.; Dass,~C.~K.;
  Sch{\"{u}}ller,~C.; Korn,~T.; Richter,~M. \latin{et~al.}  {Neutral and
  charged inter-valley biexcitons in monolayer MoSe$_{2}$}. \emph{Nat. Commun.}
  \textbf{2017}, \emph{8}, 15552\relax
\mciteBstWouldAddEndPuncttrue
\mciteSetBstMidEndSepPunct{\mcitedefaultmidpunct}
{\mcitedefaultendpunct}{\mcitedefaultseppunct}\relax
\EndOfBibitem
\bibitem[Hanbicki \latin{et~al.}(2018)Hanbicki, Chuang, Rosenberger, Hellberg,
  Sivaram, McCreary, Mazin, and Jonker]{Hanbicki2018}
Hanbicki,~A.~T.; Chuang,~H.-J.; Rosenberger,~M.~R.; Hellberg,~C.~S.;
  Sivaram,~S.~V.; McCreary,~K.~M.; Mazin,~I.~I.; Jonker,~B.~T. {Double Indirect
  Interlayer Exciton in a MoSe$_{2}$/WSe$_{2}$ van der Waals Heterostructure}.
  \emph{ACS Nano} \textbf{2018}, \emph{12}, 4719--4726\relax
\mciteBstWouldAddEndPuncttrue
\mciteSetBstMidEndSepPunct{\mcitedefaultmidpunct}
{\mcitedefaultendpunct}{\mcitedefaultseppunct}\relax
\EndOfBibitem
\bibitem[Island \latin{et~al.}(2016)Island, Kuc, Diependaal, Bratschitsch, {Van
  Der Zant}, Heine, and Castellanos-Gomez]{Island2016}
Island,~J.~O.; Kuc,~A.; Diependaal,~E.~H.; Bratschitsch,~R.; {Van Der
  Zant},~H.~S.; Heine,~T.; Castellanos-Gomez,~A. {Precise and reversible band
  gap tuning in single-layer MoSe$_{2}$ by uniaxial strain}. \emph{Nanoscale}
  \textbf{2016}, \emph{8}, 2589--2593\relax
\mciteBstWouldAddEndPuncttrue
\mciteSetBstMidEndSepPunct{\mcitedefaultmidpunct}
{\mcitedefaultendpunct}{\mcitedefaultseppunct}\relax
\EndOfBibitem
\bibitem[Kang \latin{et~al.}(2013)Kang, Tongay, Zhou, Li, and Wu]{Kang2013a}
Kang,~J.; Tongay,~S.; Zhou,~J.; Li,~J.; Wu,~J. {Band offsets and
  heterostructures of two-dimensional semiconductors}. \emph{Appl. Phys. Lett.}
  \textbf{2013}, \emph{102}, 012111\relax
\mciteBstWouldAddEndPuncttrue
\mciteSetBstMidEndSepPunct{\mcitedefaultmidpunct}
{\mcitedefaultendpunct}{\mcitedefaultseppunct}\relax
\EndOfBibitem
\bibitem[Zhang \latin{et~al.}(2019)Zhang, Zhang, Sperling, and
  Nguyen]{Zhang2019a}
Zhang,~Q.; Zhang,~S.; Sperling,~B.~A.; Nguyen,~N.~V. {Band Offset and Electron
  Affinity of Monolayer MoSe$_{2}$ by Internal Photoemission}. \emph{J.
  Electron. Mater.} \textbf{2019}, \emph{48}, 6446--6450\relax
\mciteBstWouldAddEndPuncttrue
\mciteSetBstMidEndSepPunct{\mcitedefaultmidpunct}
{\mcitedefaultendpunct}{\mcitedefaultseppunct}\relax
\EndOfBibitem
\bibitem[Shimada and Ohuchi(1994)Shimada, and Ohuchi]{Shimada1994}
Shimada,~T.; Ohuchi,~F.~S. {Work Function and Photothreshold of Layered Metal
  Dichalcogenides}. \emph{Jpn. J. Appl. Phys.} \textbf{1994}, \emph{33},
  2696\relax
\mciteBstWouldAddEndPuncttrue
\mciteSetBstMidEndSepPunct{\mcitedefaultmidpunct}
{\mcitedefaultendpunct}{\mcitedefaultseppunct}\relax
\EndOfBibitem
\bibitem[Yuan \latin{et~al.}(1985)Yuan, Pudensi, Vawter, and Merz]{Yuan1985}
Yuan,~Y.~R.; Pudensi,~M. A.~A.; Vawter,~G.~A.; Merz,~J.~L. {New
  photoluminescence effects of carrier confinement at an AlGaAs/GaAs
  heterojunction interface}. \emph{J. Appl. Phys.} \textbf{1985}, \emph{58},
  397--403\relax
\mciteBstWouldAddEndPuncttrue
\mciteSetBstMidEndSepPunct{\mcitedefaultmidpunct}
{\mcitedefaultendpunct}{\mcitedefaultseppunct}\relax
\EndOfBibitem
\bibitem[Rosenberger \latin{et~al.}(2018)Rosenberger, Chuang, McCreary,
  Hanbicki, Sivaram, and Jonker]{Rosenberger2018}
Rosenberger,~M.~R.; Chuang,~H.~J.; McCreary,~K.~M.; Hanbicki,~A.~T.;
  Sivaram,~S.~V.; Jonker,~B.~T. {Nano-"Squeegee" for the Creation of Clean 2D
  Material Interfaces}. \emph{ACS Appl. Mater. Interfaces} \textbf{2018},
  \emph{10}, 10379--10387\relax
\mciteBstWouldAddEndPuncttrue
\mciteSetBstMidEndSepPunct{\mcitedefaultmidpunct}
{\mcitedefaultendpunct}{\mcitedefaultseppunct}\relax
\EndOfBibitem
\bibitem[Barbone \latin{et~al.}(2018)Barbone, Montblanch, Kara,
  Palacios-Berraquero, Cadore, {De Fazio}, Pingault, Mostaani, Li, Chen,
  Watanabe, Taniguchi, Tongay, Wang, Ferrari, and Atat{\"{u}}re]{Barbone2018}
Barbone,~M.; Montblanch,~A. R.-P.; Kara,~D.~M.; Palacios-Berraquero,~C.;
  Cadore,~A.~R.; {De Fazio},~D.; Pingault,~B.; Mostaani,~E.; Li,~H.; Chen,~B.
  \latin{et~al.}  {Charge-tuneable biexciton complexes in monolayer WSe$_{2}$}.
  \emph{Nat. Commun.} \textbf{2018}, \emph{9}, 3721\relax
\mciteBstWouldAddEndPuncttrue
\mciteSetBstMidEndSepPunct{\mcitedefaultmidpunct}
{\mcitedefaultendpunct}{\mcitedefaultseppunct}\relax
\EndOfBibitem
\bibitem[Shibata \latin{et~al.}(2005)Shibata, Sakai, Yamada, Matsubara,
  Sakurai, Tampo, Ishizuka, Kim, and Niki]{Shibata2005}
Shibata,~H.; Sakai,~M.; Yamada,~A.; Matsubara,~K.; Sakurai,~K.; Tampo,~H.;
  Ishizuka,~S.; Kim,~K.-K.; Niki,~S. {Excitation-Power Dependence of Free
  Exciton Photoluminescence of Semiconductors}. \emph{Jpn. J. Appl. Phys.}
  \textbf{2005}, \emph{44}, 6113--6114\relax
\mciteBstWouldAddEndPuncttrue
\mciteSetBstMidEndSepPunct{\mcitedefaultmidpunct}
{\mcitedefaultendpunct}{\mcitedefaultseppunct}\relax
\EndOfBibitem
\bibitem[O'Donnell and Chen(1991)O'Donnell, and Chen]{ODonnell1991}
O'Donnell,~K.~P.; Chen,~X. {Temperature dependence of semiconductor band gaps}.
  \emph{Appl. Phys. Lett.} \textbf{1991}, \emph{58}, 2924--2926\relax
\mciteBstWouldAddEndPuncttrue
\mciteSetBstMidEndSepPunct{\mcitedefaultmidpunct}
{\mcitedefaultendpunct}{\mcitedefaultseppunct}\relax
\EndOfBibitem
\bibitem[Shibata(1998)]{Shibata1998}
Shibata,~H. {Negative Thermal Quenching Curves in Photoluminescence of Solids}.
  \emph{Jpn. J. Appl. Phys.} \textbf{1998}, \emph{37}, 550--553\relax
\mciteBstWouldAddEndPuncttrue
\mciteSetBstMidEndSepPunct{\mcitedefaultmidpunct}
{\mcitedefaultendpunct}{\mcitedefaultseppunct}\relax
\EndOfBibitem
\bibitem[Huang \latin{et~al.}(2016)Huang, Hoang, and Mikkelsen]{Huang2016}
Huang,~J.; Hoang,~T.~B.; Mikkelsen,~M.~H. {Probing the origin of excitonic
  states in monolayer WSe$_{2}$}. \emph{Sci. Rep.} \textbf{2016}, \emph{6},
  22414\relax
\mciteBstWouldAddEndPuncttrue
\mciteSetBstMidEndSepPunct{\mcitedefaultmidpunct}
{\mcitedefaultendpunct}{\mcitedefaultseppunct}\relax
\EndOfBibitem
\bibitem[Lin \latin{et~al.}(2012)Lin, Chen, Xiong, Yang, He, and Luo]{Lin2012}
Lin,~S.~S.; Chen,~B.~G.; Xiong,~W.; Yang,~Y.; He,~H.~P.; Luo,~J. {Negative
  thermal quenching of photoluminescence in zinc oxide
  nanowire-core/graphene-shell complexes}. \emph{Opt. Express} \textbf{2012},
  \emph{20}, A706\relax
\mciteBstWouldAddEndPuncttrue
\mciteSetBstMidEndSepPunct{\mcitedefaultmidpunct}
{\mcitedefaultendpunct}{\mcitedefaultseppunct}\relax
\EndOfBibitem
\bibitem[Mouri \latin{et~al.}(2017)Mouri, Zhang, Kozawa, Miyauchi, Eda, and
  Matsuda]{Mouri2017}
Mouri,~S.; Zhang,~W.; Kozawa,~D.; Miyauchi,~Y.; Eda,~G.; Matsuda,~K. {Thermal
  dissociation of inter-layer excitons in MoS$_{2}$/MoSe$_{2}$
  hetero-bilayers}. \emph{Nanoscale} \textbf{2017}, \emph{9}, 6674--6679\relax
\mciteBstWouldAddEndPuncttrue
\mciteSetBstMidEndSepPunct{\mcitedefaultmidpunct}
{\mcitedefaultendpunct}{\mcitedefaultseppunct}\relax
\EndOfBibitem
\bibitem[Chen \latin{et~al.}(2018)Chen, Goldstein, Taniguchi, Watanabe, and
  Yan]{Chen2018b}
Chen,~S.-Y.; Goldstein,~T.; Taniguchi,~T.; Watanabe,~K.; Yan,~J. {Coulomb-bound
  four- and five-particle intervalley states in an atomically-thin
  semiconductor}. \emph{Nat. Commun.} \textbf{2018}, \emph{9}, 3717\relax
\mciteBstWouldAddEndPuncttrue
\mciteSetBstMidEndSepPunct{\mcitedefaultmidpunct}
{\mcitedefaultendpunct}{\mcitedefaultseppunct}\relax
\EndOfBibitem
\bibitem[Coldren \latin{et~al.}(2012)Coldren, Corzine, and
  Ma{\v{s}}anovi{\'{c}}]{LarryA.ColdrenScottW.Corzine2012}
Coldren,~L.~A.; Corzine,~S.~W.; Ma{\v{s}}anovi{\'{c}},~M.~L. \emph{{Diode
  Lasers and Photonic Integrated Circuits}}, 2nd ed.; John Wiley \& Sons, Inc.,
  2012\relax
\mciteBstWouldAddEndPuncttrue
\mciteSetBstMidEndSepPunct{\mcitedefaultmidpunct}
{\mcitedefaultendpunct}{\mcitedefaultseppunct}\relax
\EndOfBibitem
\bibitem[Chuang(2009)]{Chuang2009}
Chuang,~S.~L. \emph{{Physics of Photonic Devices}}, 2nd ed.; Wiley, 2009\relax
\mciteBstWouldAddEndPuncttrue
\mciteSetBstMidEndSepPunct{\mcitedefaultmidpunct}
{\mcitedefaultendpunct}{\mcitedefaultseppunct}\relax
\EndOfBibitem
\bibitem[Lindemann \latin{et~al.}(2019)Lindemann, Xu, Pusch, Michalzik,
  Hofmann, {\v{Z}}uti{\'{c}}, and Gerhardt]{Lindemann2019}
Lindemann,~M.; Xu,~G.; Pusch,~T.; Michalzik,~R.; Hofmann,~M.~R.;
  {\v{Z}}uti{\'{c}},~I.; Gerhardt,~N.~C. {Ultrafast spin-lasers}. \emph{Nature}
  \textbf{2019}, \emph{568}, 212--215\relax
\mciteBstWouldAddEndPuncttrue
\mciteSetBstMidEndSepPunct{\mcitedefaultmidpunct}
{\mcitedefaultendpunct}{\mcitedefaultseppunct}\relax
\EndOfBibitem
\bibitem[Castellanos-Gomez \latin{et~al.}(2014)Castellanos-Gomez, Buscema,
  Molenaar, Singh, Janssen, van~der Zant, and Steele]{Castellanos-Gomez2014}
Castellanos-Gomez,~A.; Buscema,~M.; Molenaar,~R.; Singh,~V.; Janssen,~L.;
  van~der Zant,~H. S.~J.; Steele,~G.~A. {Deterministic transfer of
  two-dimensional materials by all-dry viscoelastic stamping}. \emph{2D Mater.}
  \textbf{2014}, \emph{1}, 011002\relax
\mciteBstWouldAddEndPuncttrue
\mciteSetBstMidEndSepPunct{\mcitedefaultmidpunct}
{\mcitedefaultendpunct}{\mcitedefaultseppunct}\relax
\EndOfBibitem
\bibitem[Singh and Nordstrom(2006)Singh, and Nordstrom]{Singh2006}
Singh,~D.~J.; Nordstrom,~L. \emph{{Planewaves, Pseudopotentials and the LAPW
  Method}}, 2nd ed.; Springer US, 2006\relax
\mciteBstWouldAddEndPuncttrue
\mciteSetBstMidEndSepPunct{\mcitedefaultmidpunct}
{\mcitedefaultendpunct}{\mcitedefaultseppunct}\relax
\EndOfBibitem
\bibitem[Blaha \latin{et~al.}(2001)Blaha, Schwarz, Madsen, Kvasnicka, Luitz,
  Laskowski, Tran, and Marks]{Blaha2001}
Blaha,~P.; Schwarz,~K.; Madsen,~G. K.~H.; Kvasnicka,~D.; Luitz,~J.;
  Laskowski,~R.; Tran,~F.; Marks,~L.~D. \emph{{WIEN2k : An augmented plane wave
  plus local orbitals program for calculating crystal properties}}; Techn.
  Universitat, 2001\relax
\mciteBstWouldAddEndPuncttrue
\mciteSetBstMidEndSepPunct{\mcitedefaultmidpunct}
{\mcitedefaultendpunct}{\mcitedefaultseppunct}\relax
\EndOfBibitem
\bibitem[Perdew \latin{et~al.}(1996)Perdew, Burke, and Ernzerhof]{Perdew1996}
Perdew,~J.~P.; Burke,~K.; Ernzerhof,~M. {Generalized gradient approximation
  made simple}. \emph{Phys. Rev. Lett.} \textbf{1996}, \emph{77},
  3865--3868\relax
\mciteBstWouldAddEndPuncttrue
\mciteSetBstMidEndSepPunct{\mcitedefaultmidpunct}
{\mcitedefaultendpunct}{\mcitedefaultseppunct}\relax
\EndOfBibitem
\bibitem[Grimme(2006)]{Allouche2012}
Grimme,~S. {Semiempirical GGA-type density functional constructed with a
  long-range dispersion correction}. \emph{J. Comput. Chem.} \textbf{2006},
  \emph{27}, 1787--1799\relax
\mciteBstWouldAddEndPuncttrue
\mciteSetBstMidEndSepPunct{\mcitedefaultmidpunct}
{\mcitedefaultendpunct}{\mcitedefaultseppunct}\relax
\EndOfBibitem
\end{mcitethebibliography}
\end{document}